%% file: main.tex
\documentclass[sigconf]{acmart}
\AtBeginDocument{%
  \providecommand\BibTeX{{%
    \normalfont B\kern-0.5em{\scshape i\kern-0.25em b}\kern-0.8em\TeX}}}

\newcommand{\quotedtext}[1]{%
   \noindent\textbf{\textit{{"#1"}}}%
}
\newcommand{\customheading}[1]{%
   \noindent\textbf{{#1}}%
}

\usepackage{enumitem}
\usepackage{float}
\usepackage{soul}
\usepackage{xcolor}
\usepackage{subcaption}

\DeclareRobustCommand{\hlcyan}[1]{{\sethlcolor{white}\hl{#1}}}

\sethlcolor{white}
\copyrightyear{2024}
\acmYear{2024}
\setcopyright{acmlicensed}\acmConference[ACE 2024]{Australian Computing Education Conference}{January 29-February 2, 2024}{Sydney, NSW, Australia}
\acmBooktitle{Australian Computing Education Conference (ACE 2024), January 29-February 2, 2024, Sydney, NSW, Australia}
\acmPrice{15.00}
\acmDOI{10.1145/3636243.3636257}
\acmISBN{979-8-4007-1619-5/24/01}
\begin{document}
\title[It's not like Jarvis, but it's pretty close!]{"It's not like Jarvis, but it's pretty close!" - Examining ChatGPT's Usage among Undergraduate Students in Computer Science}




\author{Ritvik Budhiraja\textsuperscript{$ \dag  $}}
\email{ritvik19322@iiitd.ac.in}
\affiliation{%
  \institution{IIIT Delhi}
  \city{New Delhi}
  \country{India}
}

\author{Ishika Joshi\textsuperscript{$ \dag  $}}
\email{ishika19310@iiitd.ac.in}
\affiliation{%
  \institution{IIIT Delhi}
  \city{New Delhi}
  \country{India}
  }

\author{Harshal Akolekar}
\email{harshal.akolekar@iitj.ac.in}
\affiliation{%
  \institution{Dept of Mech. Eng. \& School of AIDE}
  \city{IIT Jodhpur, Jodhpur}
  \country{India}
}

\author{Jagat Sesh Challa}
\email{jagatsesh@pilani.bits-pilani.ac.in}
\affiliation{%
  \institution{BITS Pilani}
  \city{Pilani}
  \country{India}
}

\author{Dhruv Kumar}
\email{dhruv.kumar@iiitd.ac.in}
\affiliation{%
  \institution{IIIT Delhi}
  \city{New Delhi}
  \country{India}
}

\titlenote{Jarvis : https://en.wikipedia.org/wiki/J.A.R.V.I.S.\\
\textsuperscript{$ \dag  $}Equal Contribution}

\renewcommand{\shortauthors}{Budhiraja, Joshi, et al.}

\begin{abstract}
\input{files/00-abstract}
\end{abstract}





\maketitle

\section{Introduction}
\input{files/01-introduction}

\section{Related Work}
\input{files/02-related_work}
\section{Methodology}
\input{files/03-methodology}
\section{Quantitative Evaluation}
\input{files/04x-Quantitative_Evaluation}
\section{Qualitative Evaluation}
\input{files/04-evaluation}
\vspace{-1em}
\section{Discussion}
\input{files/05-discussion}
\section{Conclusion}
\input{files/06-conclusion}
\begin{acks}
\input{files/061-acknowledgements}
\end{acks}
\bibliographystyle{ACM-Reference-Format}
\bibliography{chatgpt-1}

\appendix
\input{files/07-appendix}

\end{document}

%% file: files/00-abstract.tex
Large language models (LLMs) such as ChatGPT and Google Bard have garnered significant attention in the academic community. Previous research has evaluated these LLMs for various applications such as generating programming exercises and solutions. However, these evaluations have predominantly been conducted by instructors and researchers, not considering the actual usage of LLMs by students. This study adopts a student-first approach to comprehensively understand how undergraduate computer science students utilize ChatGPT, a popular LLM, released by OpenAI. We employ a combination of student surveys and interviews to obtain valuable insights into the benefits, challenges, and suggested improvements related to ChatGPT. Our findings suggest that a majority of students (over 57\%) have a convincingly positive outlook towards adopting ChatGPT as an aid in coursework-related tasks. However, our research also highlights various challenges that must be resolved for long-term acceptance of ChatGPT amongst students. The findings from this investigation have broader implications and may be applicable to other LLMs and their role in computing education.

%% file: files/01-introduction.tex
Large Language Models (LLMs) \cite{floridi2020gpt} have garnered significant attention recently due to their remarkable capacity to perform complex tasks. These models are integral to the Generative AI paradigm \cite{gen_ai_web_link}, as they not only extract information from existing data but also generate novel content based on it. Prominent examples of LLMs include ChatGPT \cite{ChatGPT} and Codex \cite{OpenAI_Codex}, developed by OpenAI, Bard \cite{Google_Bard} by Google, and AlphaCode \cite{alphacode} by DeepMind. Within the academic community also, there have been extensive discussions about the challenges, risks, and opportunities involved in their utilization from both student and instructor perspectives \cite{becker2023ProsAndCons, Denny2023CopilotCS1, Finnie-Ansley2022CS1}.

Numerous research studies have been conducted in the computing education community concerning Large Language Models (LLMs) \cite{becker2023ProsAndCons, Malinka2023Security, Daun2023Software, Denny2023CopilotCS1, Finnie-Ansley2022CS1, wermelinger2023Copilot, Savelka2023MCQAndCode, Reeves2023Parsons, finnie-ansley2023CodexCS2, Ouh2023Java, Cipriano2023GPT-3OOP, sarsa2022AutoGenerate, Leinonen2023CodeExplanation, Leinonen2023ExplainError, MacNeil2023CodeExplain, Balse2023Feedback}. Educators have assessed the performance of these LLMs in generating precise solutions to questions posed to students both within and beyond the classroom setting \cite{Denny2023CopilotCS1, Finnie-Ansley2022CS1, wermelinger2023Copilot, Reeves2023Parsons, Savelka2023MCQAndCode, Ouh2023Java, Cipriano2023GPT-3OOP, Daun2023Software, Malinka2023Security}. Other studies have looked at using LLMs to generate fresh programming exercises and code explanations \cite{sarsa2022AutoGenerate, wermelinger2023Copilot, Leinonen2023CodeExplanation, MacNeil2023CodeExplain}, provide suggestive fixes to code errors \cite{Leinonen2023ExplainError} and generate personalized feedback for students based on their code submissions \cite{Balse2023Feedback}.

However, there remains a lack of student-centred research that studies the perceptions and current usage of LLMs by students in academic contexts. This paper addresses this gap by adopting a student-first approach through surveys and interviews to examine the real-world usage of LLMs amongst undergraduate computer science (CS) students. Our study makes an essential contribution to the existing body of literature. Firstly, it provides suggestions and recommendations to instructors and educators, enabling them to adapt and refine their instructional approaches. Secondly, it can aid researchers working in the space of technology-assisted education in building their research on the insights suggested in this study and shape their future research directions in this domain. Thirdly, a thorough comprehension of the usage patterns and strategies can inform the design of appropriate interfaces to enhance the utilization of LLMs among students.

Our paper employs a mixed-methods approach \cite{mixed_methods}, involving qualitative and quantitative assessment. We gather 480 survey responses and hold 17 interviews with CS undergraduates in Indian universities. The surveys and interviews focus on addressing the following research questions:
\begin{itemize}
    \item \textbf{RQ1:} What are the usage patterns of ChatGPT among students and their perceived benefits?
    \item \textbf{RQ2:} What are the challenges faced by students in utilizing ChatGPT?
    \item \textbf{RQ3:} Do the students have any suggestions and recommendations for improving their usage experience of ChatGPT?
\end{itemize}

Our findings indicate that over 57\% of students view ChatGPT positively for coursework-related queries. Yet, challenges related to usability, reliability, learning, and ethics, need resolution for wider student acceptance. Although our investigation centers on ChatGPT, we firmly believe that the insights and conclusions derived from this study are also relevant to other LLMs. To the best of our knowledge, this comprehensive study represents the first detailed exploration of the practical usage of LLMs among computer science undergraduates.

%% file: files/02-related_work.tex
LLM \cite{floridi2020gpt} tools offer diverse benefits to undergraduate computer science students. One of the early studies by Becker et al. \cite{becker2023ProsAndCons} examined various challenges and opportunities associated with the utilization of AI code generation tools including OpenAI Codex \cite{OpenAI_Codex}, DeepMind AlphaCode \cite{alphacode}, and Amazon CodeWhisperer \cite{codeWhisperer}. This paper discusses that LLM tools can be useful for a wide variety of purposes such as generating multiple code solutions for a particular problem, generating quality learning material for students, generating starter code for students to get started, and debugging code. However, there are a number of issues related to ethics, bias, and security which need to be addressed for wider adoption of these tools.
Similar challenges and opportunities have also been addressed in other studies \cite{Denny2023CopilotCS1, Malinka2023Security, Daun2023Software}. 

Numerous research studies have been dedicated to assessing the accuracy of LLMs, such as OpenAI Codex \cite{OpenAI_Codex}, GPT-3 \cite{gpt3}, ChatGPT (GPT-3.5 and GPT-4) \cite{ChatGPT}, in generating solutions for programming assignments across various computer science courses, including CS1 \cite{Denny2023CopilotCS1, Finnie-Ansley2022CS1, wermelinger2023Copilot, Savelka2023MCQAndCode, Reeves2023Parsons}, CS2 \cite{finnie-ansley2023CodexCS2, Savelka2023MCQAndCode}, object-oriented programming \cite{Ouh2023Java, Cipriano2023GPT-3OOP}, software engineering \cite{Daun2023Software}, and computer security \cite{Malinka2023Security}. 
These studies demonstrate that LLMs can solve a significant portion of programming questions effectively, influenced by task complexity and prompt quality. While they generate accurate solutions, students must still cultivate skills like algorithmic thinking, program comprehension, debugging, and communication.

Leinonen et al. \cite{Leinonen2023ExplainError} assessed OpenAI Codex's ability to explain code error messages and the effectiveness of its suggested fixes. The authors confirm LLMs indeed improve error messages for novice programmers, aiding them in debugging. LLM-generated messages can even surpass original ones in interpretability and actionability.

Balse et al. \cite{Balse2023Feedback} explored the potential of GPT-3 in delivering detailed and personalized feedback to students for programming assessments. The study analyzed GPT-3 model's performance for evaluating code correctness, accurately critiquing code, and suggesting appropriate changes to the code. The analysis showed that the GPT-3 model cannot be directly used for providing feedback to the students as it's accuracy is highly variable, but it may be useful under supervision. 

Sarsa et al. \cite{sarsa2022AutoGenerate} explored using OpenAI Codex for automatically creating programming exercises and explanations, providing instructors with innovative assignment options. The study indicated that while most generated content is fresh and logical, oversight is necessary to ensure content quality before student delivery.
Several studies have analyzed the LLM models' ability to generate code explanations, contrasting the quality of these explanations with those provided by students \cite{Leinonen2023CodeExplanation, sarsa2022AutoGenerate, wermelinger2023Copilot, MacNeil2023CodeExplain}. 


Several user studies have concentrated on the examination of students' perceptions and experiences. Shoufan's study \cite{Shoufan23studentperception}, for instance, involved a user study comprising 56 senior students enrolled in a core course on embedded systems during the Spring of 2023 in a computer engineering program at a university in the United Arab Emirates. The study employed one open-ended question and 27 survey questions to gather user data. In contrast to Shoufan's work \cite{Shoufan23studentperception}, our research encompasses a more extensive and diverse sample of undergraduate students from India. Furthermore, our study incorporates both semi-structured interviews and surveys, enabling us to gain deeper insights. Similarly, Firat's research \cite{Mehmet23studentperception} analyzed the perception and experience of scholars and Ph.D. students across multiple countries, while our investigation focuses specifically on analyzing these aspects for undergraduate students.

%% file: files/03-methodology.tex
    \subsection{Research Design} 
    We adopted a mixed-methods \cite{mixed_methods} research design for this study to investigate the effects of ChatGPT on undergraduate computer science (CS) education. A mixed-methods design allowed us to integrate both qualitative and quantitative approaches, providing a comprehensive and nuanced understanding of the research problem. The participants for this study span over 10 Indian Engineering Universities. A survey, via Google Forms, was circulated among undergraduate CS students in three of the ten universities offering CS courses(Indraprastha Institute of Information Technology, Delhi[IIITD], Birla Institute of Technology And Science, Pilani [BITS Pilani], Indian Institute of Technology, Jodhpur[IITJ]). The purpose of this survey was to collect quantitative data on usage patterns of ChatGPT among these students. \hl{We shared the survey with approximately 3500 students and received responses from 480 students}. \hl{We have also carried out interviews with 17 (9F/8M) undergraduate CS students spread across different academic years in eight universities out of the ten universities mentioned above} (refer to table \ref{tab:interview_demographics}). The interview participants were recruited through snowball sampling \cite{Snowball}. The interviews were conducted over Google Meet and the audio recordings of these interviews were further analysed. The interviewees provided both written and verbal consent for the interview and the recordings.

    \subsection{Data Collection and Analysis} 

 \customheading{Survey.}
    Keeping standard survey design guidelines in mind, a short and engaging online survey was designed to capture CS students' usage and experiences with ChatGPT. The survey explored multiple aspects of ChatGPT in an academic setting, such as frequency of use (two questions) to capture the academic usage of ChatGPT, use cases (two questions) to observe the scenarios where CS students rely on ChatGPT, and overall advantages, disadvantages, and opinions on the use of ChatGPT as a tool in the domain of computer science education (six questions). Such themes were covered in order to enable us to establish the background of common CS student perceptions, challenges, and expectations from ChatGPT. The estimated time of completion for the survey was 3-5 minutes, which assisted in preventing students from avoiding the survey due to an excess of questions. The survey utilized single-select, and multi-select options along with textual response fields wherever required, allowing the collection of both qualitative and quantitative data. 
    The survey was circulated amongst university students using university student email lists. 
    Responses to the survey were analyzed and utilized further in the study to gain insights and in the framing of the interview questions. \hl{The complete list of questions asked in the survey is presented in Appendix} \ref{sec:survey_questionnaire}.
    
    \customheading{Interviews.}
The interviews aimed to comprehend CS students' academic experiences with ChatGPT in various academic scenarios such as coding, assignments, content generation, and more. Building upon such scenarios, the interview questions aimed at generating deeper insights into a student's undergraduate computer science LLM journey, including motivations, habits, benefits, challenges, perceptions, biases, and assumptions. Focusing on ChatGPT and CS, participants were encouraged to share insights into how ChatGPT supported their learning and workflow processes for their CS use cases, highlight the challenges they faced while using ChatGPT and their solutions to tackle said challenges, and were probed to share their underlying beliefs and opinions with a focus on whether they perceived ChatGPT as a threat or assistance in their CS learning experiences. Semi-structured interviews were conducted over Google Meet, which were built upon the research questions established and were refined using the initial findings from the survey. The interview was designed to maintain the survey's theme around academic Computer Science, while enabling deeper insights due to its personalized, one-on-one nature, overcoming the limitations of the brief survey format. A total of six questions were included in the structure of the interview. Certain protocols and guidelines were established, such as the overall process to follow, what is the aim of each question (eg: awareness, familiarity, shortcomings), and what probing questions are to be used if needed, offering flexibility for the research team to delve deeper. This created a dynamic interview flow that accommodated insights while maintaining overall structure across participants. The interview participants (nine female, eight male) were recruited across universities and all academic years in order to capture a broader spectrum of opinions and perceptions. No academic performance criteria were employed other than the student being enrolled in an ongoing undergraduate CS program. \hl{The list of questions used for the interviews conducted by the research team is shared in Appendix} \ref{sec:interview_questionnaire}.

    \customheading{Data Analysis.}
    The interviews were transcribed verbatim by the research team. This was followed by a Thematic Analysis (TA) \cite{braun_using_2006} on the collected data. The survey responses were coded and grouped under themes, while a three-layer TA approach was followed for the interview transcripts. This involved an initial semantic coding \cite{braun_using_2006} of the transcripts, grouping of these codes into a set of intermediate themes and a latent coding \cite{braun_using_2006} of the initial themes to reveal the final set of themes. The final themes from the surveys and interviews informed our findings and discussion. 

   \subsection{Ethical Considerations}
   In conducting this study, we undertook multiple ethical considerations to ensure transparency and maintain participant privacy and well-being.  All the research materials and protocols for this study were reviewed and approved
by our university's Institutional Review Board (IRB). Prior to participation, all candidates involved in interviews were provided with consent forms, which outlined the purpose of the study, the voluntary nature of their involvement, and the assurance of anonymity and confidentiality. We also recognize the importance of capturing the diverse perspectives and experiences of students regarding ChatGPT's impact on their education. Hence, we made efforts to ensure diversity and inclusion by inviting participants from different academic levels and genders. Participants' consent was taken to record the interviews.

   \subsection{Limitations}
   The study is designed in a way that no part includes any university-specific component. But due to limited resources and the time-bound nature of the study, it could include students from only a handful number of universities. \hl{The survey could be shared with students in a university only after we have received the required approval for bulk emailing in that university. Hence, we were able to share the survey only in three universities. As the number of conducted interviews was limited, we were able to recruit interview participants from eight universities using snowball sampling.} This may introduce unacknowledged biases pertaining to the academic capacities and practices of the included universities despite having a diverse and abundant pool of participants.

   \hl{Moreover, all student interviews were exclusively conducted by the student members of our research team, without any involvement from faculty or staff. Our team explicitly assured the students that the interview discussions would serve solely for research purposes and would not result in any disciplinary actions concerning any improper use of ChatGPT by the students. This approach was implemented to assure an environment of trust and encourage candid sharing of views without the apprehension of facing any penalties. However, it is plausible that some students may have chosen not to fully disclose their thoughts and experiences due to concerns about potential repercussions for the inappropriate utilization of ChatGPT.} 

   Due to the exploratory nature of the study, the research material like surveys and interview questionnaires are not validated by prior literature on similar research. Moreover, the nature of the surveys and questionnaires can be further studied and modified to get a deeper understanding of user behaviours and perceptions.


%% file: files/04x-Quantitative_Evaluation.tex

\begin{figure*}
    \centering
    \begin{subfigure}{0.45\textwidth}
        \includegraphics[width=0.95\textwidth]{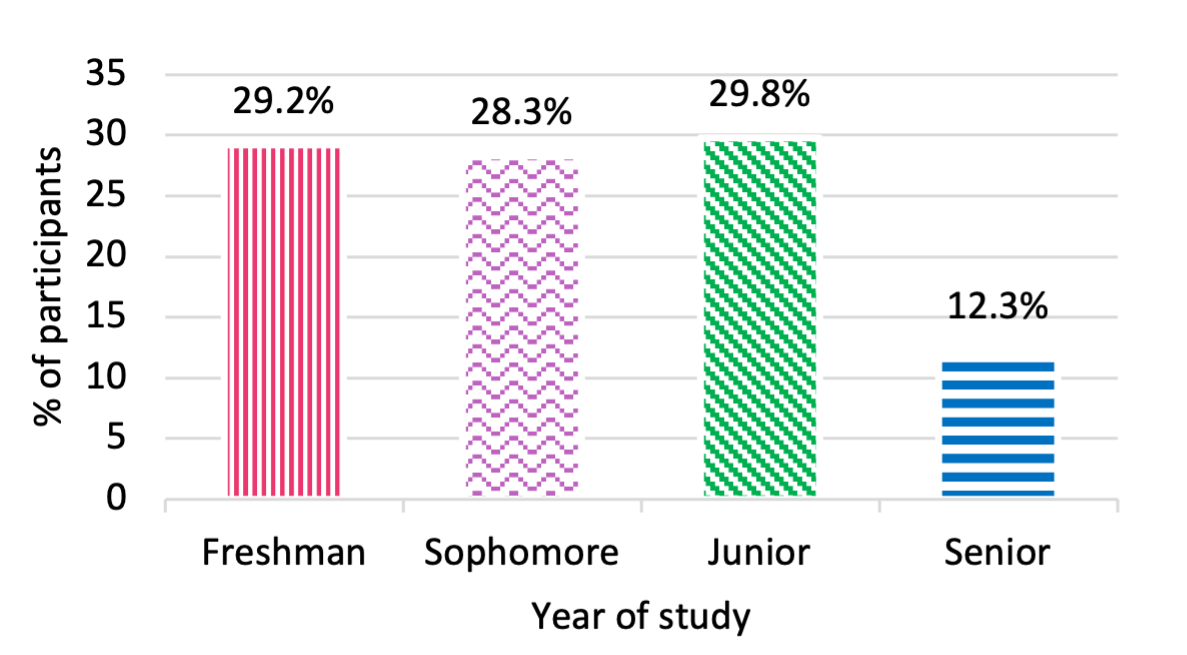}
        \caption{Year-wise distribution of survey participants}
        \label{fig:year_distr}
    \end{subfigure}
    \begin{subfigure}{0.45\textwidth}
        \includegraphics[width=0.95\textwidth]{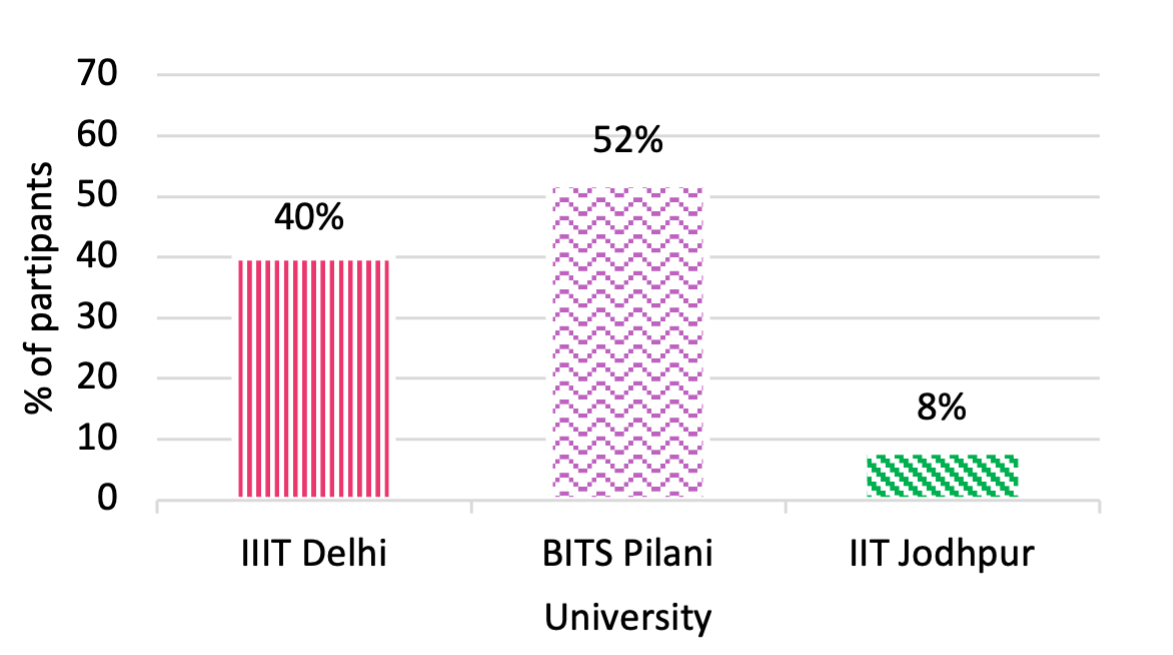}
        \caption{University-wise distribution of survey participants}
        \label{fig:univ_distr}
    \end{subfigure}
    
    \begin{subfigure}{0.45\textwidth}
        \includegraphics[width=0.95\textwidth]{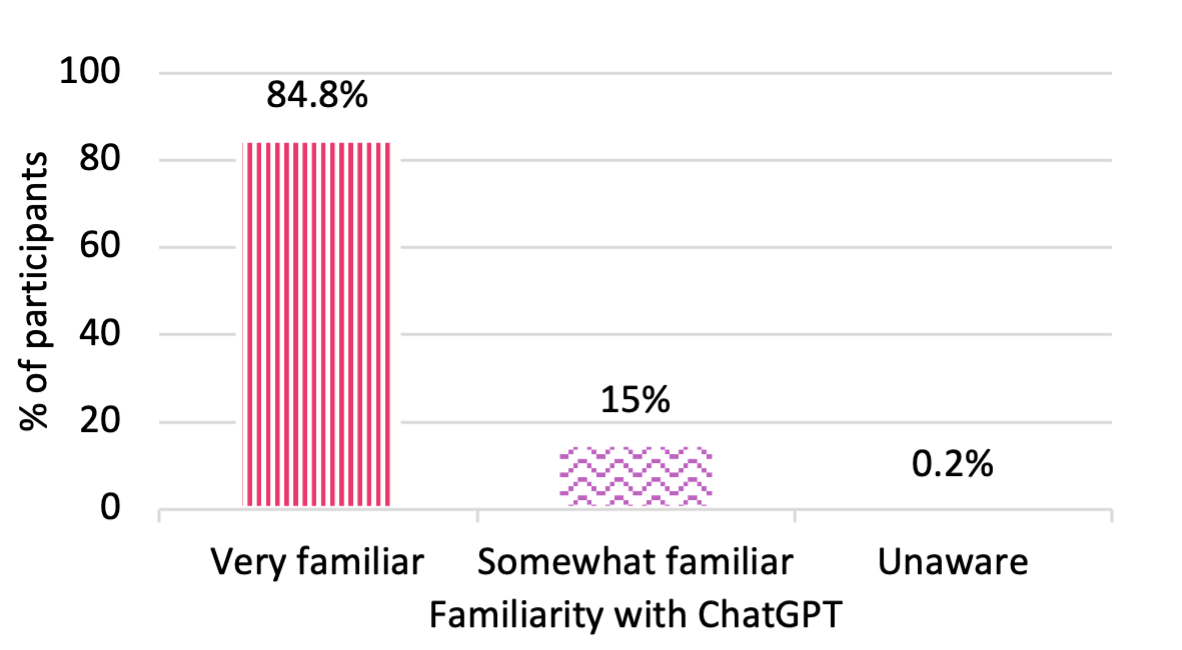}
        \caption{Familiarity with using ChatGPT by the survey participants}
        \label{fig:famil}
    \end{subfigure}
    \begin{subfigure}{0.45\textwidth}
        \includegraphics[width=0.95\textwidth]{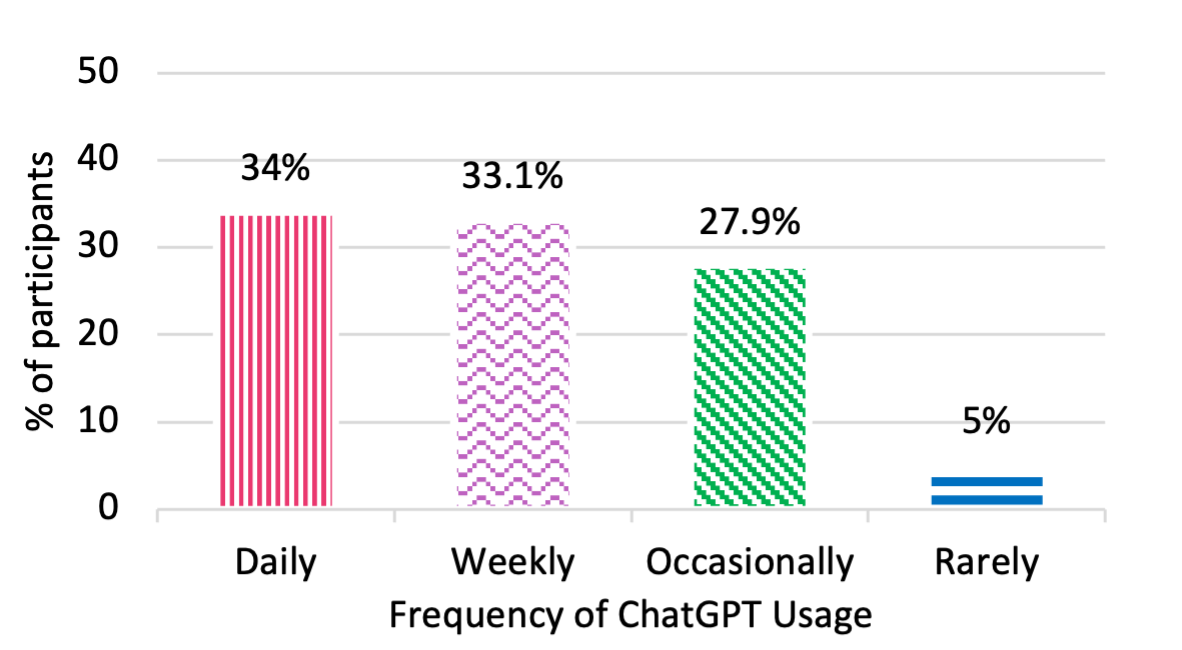}
        \caption{Usage frequency of ChatGPT by the survey participants}
        \label{fig:usage_freq}
    \end{subfigure}
    
    \begin{subfigure}{0.45\textwidth}
        \includegraphics[width=0.95\textwidth]{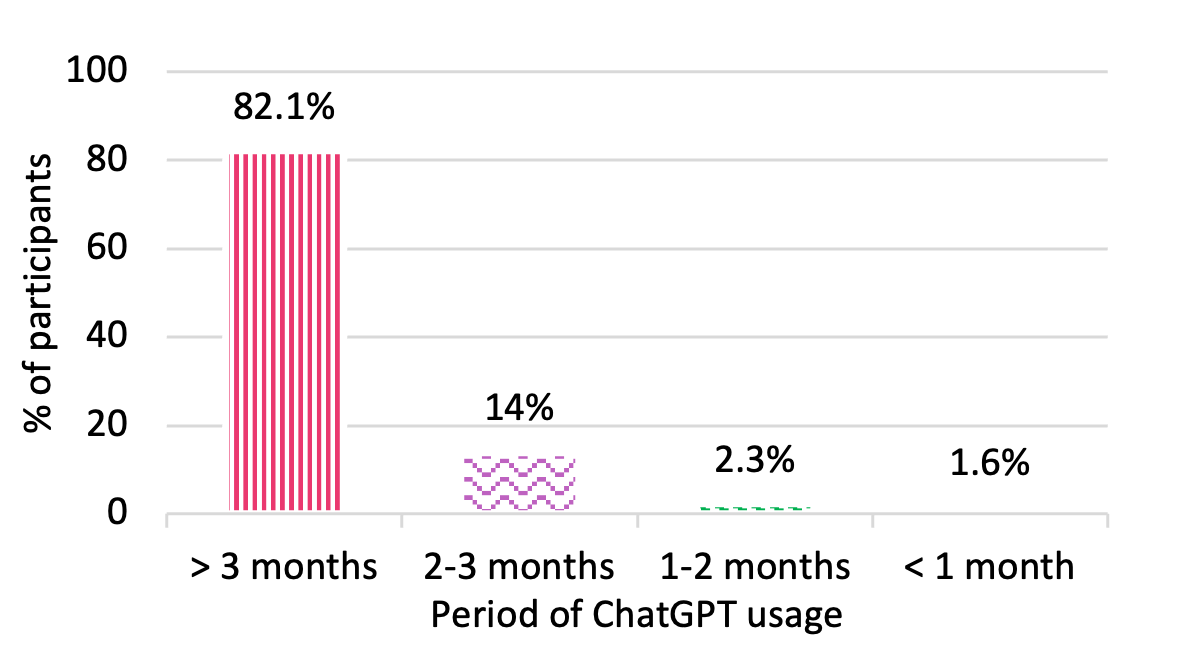}
        \caption{Period of usage of ChatGPT by the survey participants}
        \label{fig:period_usage}
    \end{subfigure}
    \begin{subfigure}{0.45\textwidth}
        \includegraphics[width=0.95\textwidth]{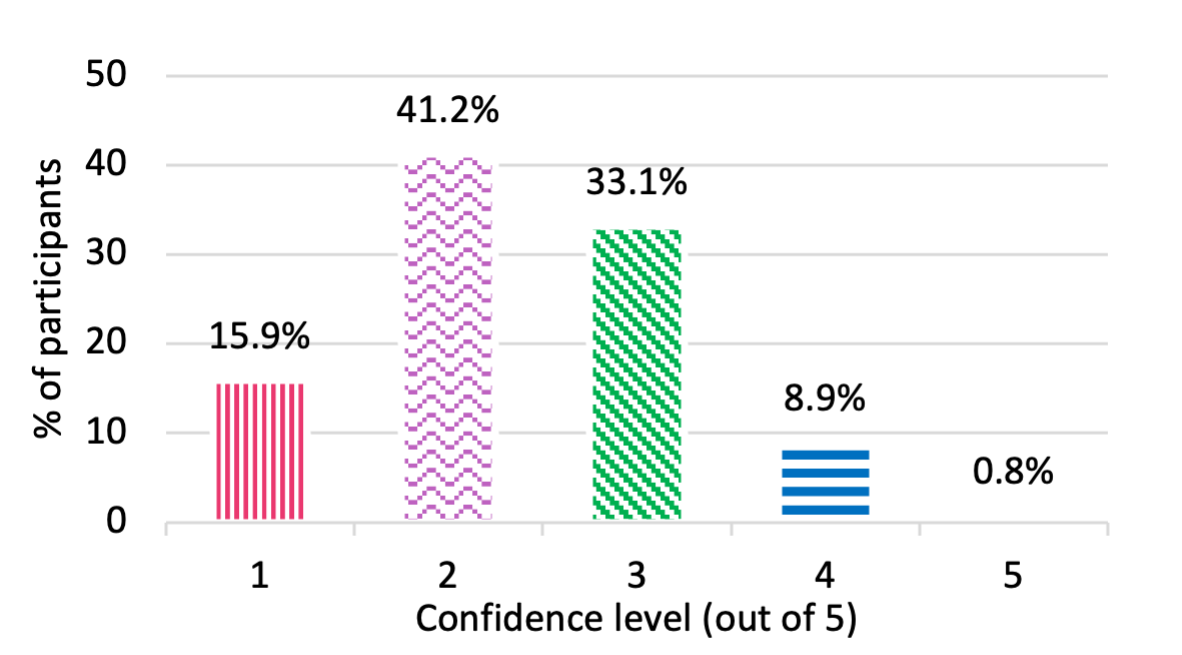}
        \caption{Level of confidence in usage of ChatGPT by the survey participants for course-work related queries}
        \label{fig:conf_course_work}
    \end{subfigure}
    
    \begin{subfigure}{0.45\textwidth}
        \includegraphics[width=0.95\textwidth]{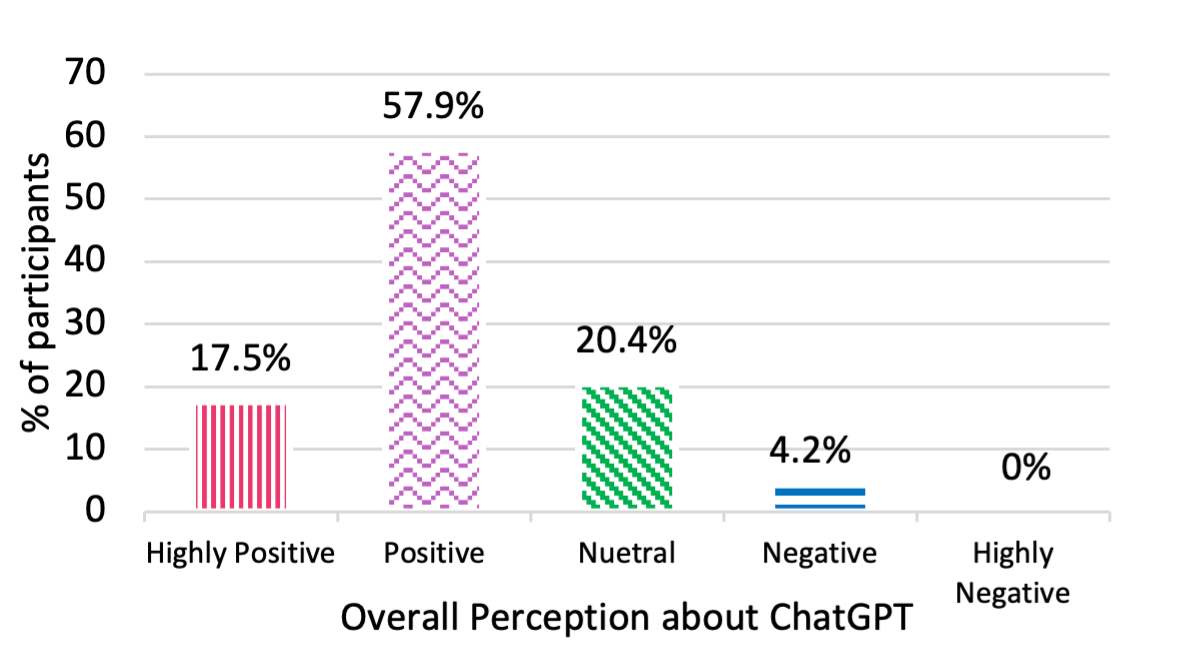}
        \caption{Overall perception of ChatGPT as an educational learning tool}
        \label{fig:overall_perc}
    \end{subfigure}
    \begin{subfigure}{0.45\textwidth}
        \includegraphics[width=0.95\textwidth]{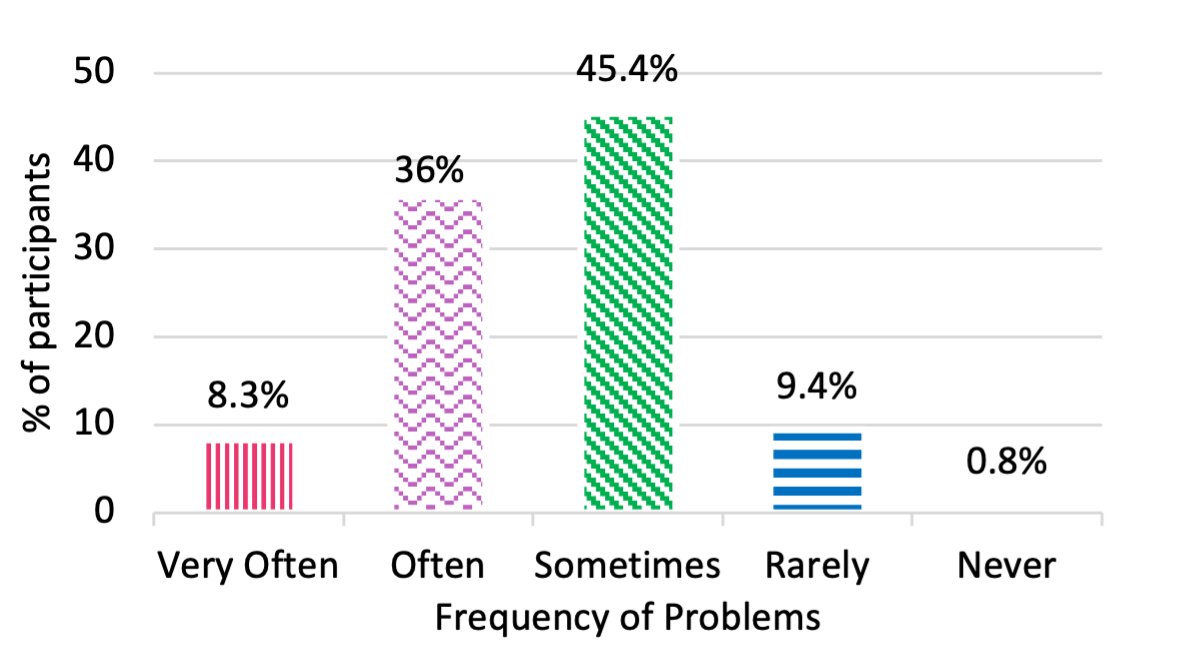}
        \caption{Frequency of response-related problems associated with ChatGPT as experienced by students}
        \label{fig:freq_response}
    \end{subfigure}
    \caption{Data visualizations for the quantitative results of the survey responses}
\end{figure*}

Using the survey formulated for the study, we were able to collect a total of 480 responses from Undergraduate Computer Science students across three universities. A typical undergraduate CS degree is a 4-year program. Figure \ref{fig:year_distr} shows the year-wise distribution of survey respondents. 29.2\% of the respondents were from the freshman year, 28.3\% from sophomore year, 29.8\% from junior year, and 12.3\% from senior year. Figure \ref{fig:univ_distr} shows the university-wise distribution of the survey participants, wherein 40\% were from University A, 52\% from University B, and 8\% from University C.

Figure \ref{fig:famil} indicates the participants' level of familiarity with using ChatGPT. It shows that 84.8\% of the respondents indicated being "very familiar" with ChatGPT, while 15.0\% of the participants noted being "somewhat familiar". An intriguing observation is that only 0.2\% of the surveyed students reported being unaware of ChatGPT, making it evident that ChatGPT has gained significant recognition within the undergraduate computer science community. Figure \ref{fig:usage_freq} shows that the usage frequency of ChatGPT amongst the CS student population follows a roughly equal distribution of daily, weekly, and occasional users. 34\% of the students were found to use ChatGPT daily, 33.1\% students used it weekly, and 27.9\% used it occasionally. 5\% of the survey population rarely used ChatGPT.
Figure \ref{fig:period_usage} shows that a majority of the participants have been long-term users of ChatGPT, with 82.1\% students reporting that they have been using it for over 3 months, 14\% being 2-3-month-old users, 2.3\% being 1-2-month-old users, and 1.6\% being new on-boarders with less than a month of use. This highlights the early adoption of the service among undergraduate CS students.

The use cases of ChatGPT in academic CS settings span multiple scenarios as reported by our participants. The survey scoped these scenarios using multiple-select questions. Table \ref{table:use_cases} depicts that the most frequent use-cases involve students using ChatGPT for gathering information (89.2\%), summarizing learning content (76.7\%),  assistance in coding (74.4\%), assistance in generating content such as academic emails and statements (68.7\%), and assistance in essay writing as required by courses(64\%). Other uses revolved around getting feedback and review (28.6\%) and testing responses of the AI (51.5\%). Individual responses were recorded using ChatGPT for academic dishonesty, as a tutor, preparing for vivas, and assistance in creative tasks, while some students reported no use of the tool due to a lack of knowledge of how to use it. These highlight how CS students have been using ChatGPT in their day-to-day academic tasks. Figure \ref{fig:conf_course_work} evaluates ChatGPT as an aid for CS coursework-related queries, with 1 being the lowest and 5 being the highest confidence level. 15.9\% students reported a confidence level of 5 and 41.2\% reported a confidence level of 4, showing that over 57\% of the population is convincingly positive. 33.1\% reported a confidence level of 3, while level 2 and level 1 were reported by 8.9\% and 0.8\% students respectively.

\begin{table}[t]
	\small
	\vspace{-1em}
    
	\begin{tabular}{|p{5cm}|p{2cm}| } 
		\hline
		\textbf{Use-Cases} & \textbf{\% of Students}\\
		\hline
		Gathering information & 82.92 \\ 
		\hline
        Summarizing content & 76.67 \\ 
		\hline
	  Assistance in essay writing & 59.79 \\
        \hline
        Assistance in coding & 74.38 \\
        \hline
        Assistance in generating content (emails, statements, etc.) & 68.75 \\
        \hline
        Assistance in assignments & 62.29 \\
        \hline
        Feedback and review & 28.75 \\
        \hline
        Testing out the AI responses & 51.46 \\
        \hline
        Other & 1.88 \\
        \hline
	\end{tabular}
    \vspace{1em}
	\caption{\textbf{Different ways of ChatGPT usage by Students}}
	\label{table:use_cases}
\end{table}

\begin{table}[t]
	\small
	\vspace{-1em}
    
	\begin{tabular}{|p{5cm}|p{2cm}| } 
		\hline
		\textbf{Advantages} & \textbf{\% of Students}\\
		\hline
		Personalized Learning & 61.67 \\ 
		\hline
        Immediate Feedback & 69.17 \\ 
		\hline
	  On-Demand Tutoring & 54.79 \\
        \hline
        Engaging and Interactive Learning & 43.33 \\
        \hline
        Access to More Information & 62.08 \\
        \hline
        Continuous Assessment & 29.58 \\
        \hline
        None & 2.29 \\
        \hline
	\end{tabular}
        \vspace{1em}
	\caption{\textbf{Advantages of ChatGPT as perceived by Students}}
	\label{table:advantages}
\end{table}

\begin{table}[t]
	\small
	\vspace{-1em}
    
	\begin{tabular}{|p{5cm}|p{2cm}| } 
		\hline
		\textbf{Challenges} & \textbf{\% of Students}\\
		\hline
		Incorrect and Unreliable answers & 74.79 \\ 
		\hline
        No way to verify the correctness & 69.38 \\ 
		\hline
	  None & 3.54 \\
        \hline
        I do not know the sources of information & 51.04 \\
        \hline
        Lack of trust & 39.17 \\
        \hline
        Other & 4.17 \\
        \hline
        Interface is inefficient & 9.17 \\
        \hline
        It is not conversational enough & 10.21 \\
        \hline
	\end{tabular}
    \vspace{1em}
	\caption{\textbf{Challenges of ChatGPT as perceived by Students}}
	\label{table:challenges}
	\vspace{-2em}
\end{table}

\begin{table}[htbp]
    \small
  \centering
  
    \begin{tabular}{|l|l|l|l|}
    \hline
    \textbf{ID} & \textbf{Gender} & \textbf{Year of Study} & \textbf{University} \\
    \hline
    P1    & Male  & Freshman & IIT-BHU  \\
    \hline
    P2    & Male  & Freshman & VIT \\
    \hline
    P3    & Female & Freshman & IIIT Delhi \\
    \hline
    P4    & Female & Freshman & IIIT Delhi \\
    \hline
    P5    & Female & Freshman & IGDTUW \\
    \hline
    P6    & Male  & Freshman & IIIT Hyderabad\\
    \hline
    P7    & Male  & Sophomore & VIT \\
    \hline
    P8    & Male  & Sophomore & DTU \\
    \hline
    P9    & Male  & Sophomore & IIIT Delhi \\
    \hline
    P10   & Female & Sophomore & VIT \\
    \hline
    P11   & Female & Sophomore & IIT Delhi \\
    \hline
    P12   & Female & Junior & IIIT Delhi \\
    \hline
    P13   & Female & Junior & New Era Academy \\
    \hline
    P14   & Female & Senior & IIT Delhi \\
    \hline
    P15   & Male  & Senior & IIIT Delhi \\
    \hline
    P16   & Male  & Senior & IIIT Delhi \\
    \hline
    P17   & Female & Senior & IIIT Delhi \\
    \hline
    \end{tabular}%
    \vspace{1em}
    \caption{Demographic information of Interview Participants}
  \label{tab:interview_demographics}
  \vspace{-2em}
\end{table}%

Figure \ref{fig:overall_perc} presents the overall perception of ChatGPT as a learning and education tool. 17.5\% of students are highly positive about their overall perception of ChatGPT as a learning and education tool, while 57.9\% are positive. 20.4\% reported being neutral, 4.2\% negative and 0\% highly negative. Table \ref{table:advantages}  summarizes the advantages of ChatGPT as perceived by students. They include - Immediate feedback e.g. to code, explanations, and self-written answers (69.17\%), access to more information (66.9\%), personalized learning of concepts (61.67\%), and on-demand tutoring (54.79\%). Table \ref{table:challenges} lists various issues faced by students in their usage of ChatGPT for CS course requirements. The three most frequent issues are incorrect and unreliable responses (74.8\%), lack of ways to verify the responses generated by it (69.4\%), and lack of sources for the responses provided by ChatGPT (51.0\%). Figure \ref{fig:freq_response} shows how often students run into problems while using ChatGPT. 8.3\% of students reported having run into such response-related problems very often, 36\% often, and 45.4\% sometimes. 9.4\% of students report rarely running into problems. It is worth noticing that 0.8\% have never faced such issues.



%% file: files/04-evaluation.tex
We performed a three-layer thematic analysis\cite{braun_using_2006} on the qualitative data collected from the interviews and survey as explained in the methodology. This data was collated, merged and finally categorized into three overarching themes - \textit{Usage Patterns and Benefits, Challenges, and Perceptions and Recommendations}. This enabled us to categorize and report our findings in a structured manner encompassing insights covering different aspects of the experience of using ChatGPT.

Table \ref{tab:interview_demographics} depicts the demographics of the participants of the interviews. The participants were taken from 8 different universities. Out of 17 participants, 8 were male and 9 were female. 6 participants were freshmen, 5 participants were sophomores, 2 participants were juniors and 3 were seniors. This representation is influenced by the varying availability of students across different academic years owing to commitments such as internship drive preparations and placement preparations.
\subsection{Usage Patterns and Benefits of ChatGPT}
\customheading{Usage Patterns adopted by Undergraduate Computer Science students.} Our findings show that the growing discussions and dialogue about ChatGPT was the starting point for most of our participants to try exploring this tool. Various participants talked about using it first just to discover its general and CS-specific capabilities. This was followed by trying it out for academic use cases such as assignments and take-home labs and eventually making regular use of it. Participants talked about trying it for writing codes or solving questions for their academic assignments to scope its capabilities. A participant mentioned that ChatGPT's conversational nature helped them to review and reinforce concepts and engage in discussions with it. A participant mentioned -

\quotedtext{\hlcyan{I've been using ChatGPT for about seven or eight months now. I heard about it from a fellow batchmate [...] when I was stuck on an assignment and then he recommended me to use ChatGPT for it. Initially, I used it rarely, but lately, it has become quite a handy tool.}} -[P14]

Participants talked about depending on ChatGPT for their \hlcyan{academic assignments and preparation for course tests and exams}. They mentioned that their usage would increase depending on the workload they had. Participants relied on ChatGPT when they had time constraints for assignments. They utilized it for tasks such as \hlcyan{information gathering, code generation, error identification, and more}. They broke down problems related to code and concepts, inputting them into ChatGPT while cross-checking outputs with their domain knowledge.

\quotedtext{\hlcyan{Earlier when I would have to spend 2 hours to even find a basic approach for a problem, ChatGPT would give that to me in 2 minutes, and from there I can build on that.}} -[P10]

A recurring usage pattern was observed: users sought answers from ChatGPT, cross-checked with peers or sources like Google Search, corrected errors with ChatGPT, and extracted accurate responses. \hlcyan{For code, they tested the generated code, sought error fixes from ChatGPT, and retrieved corrected versions}. Dependence on ChatGPT was observed to be very prominent, however, it has reduced as users started to recognize its current technical limitations and challenges. As a participant mentioned -

\quotedtext{\hlcyan{It significantly reduces the time I spend searching for solutions to
problems. It is like, you know, it points me in the right direction, and I can Google the rest up.}} -[P7]

\customheading{Emerging Benefits of using ChatGPT for Academic Tasks.} Our findings reveal ChatGPT's potential to support ideation among undergraduates. It is seen as a valuable tool for \hlcyan{initiating problem-solving} as it offers vast knowledge for \hlcyan{idea generation}. It was referred to as a comprehensive search engine that aids learning across various domains. The conversational aspect of it also eases onboarding and offers personalized insights. This helps with brainstorming and exploring unfamiliar CS domains. A participant mentioned -

\quotedtext{\hlcyan{It has become an indispensable tool that I rely on regularly to gather information,  brainstorm ideas, get the algorithm, and debug the code.}} -[P6]

While talking about use cases specific to computer science coursework, our participants talked about ChatGPT being a handy tool to quickly understand small details in \hlcyan{documentation and language semantics}. More so, ChatGPT could serve as an assistant to \hlcyan{debug coded algorithms} and learn more about them using its conversational nature. A participant mentioned how ChatGPT is better than other traditional online sources to assist in programming, such as Stack Overflow \cite{stack_overflow}, because one would need to recheck its generated answer using their own understanding and converse back and forth with ChatGPT, which cannot happen in other online resources where students directly copy-paste the answers they find.

\quotedtext{It is good for debugging the code someone has written because otherwise people would anyway go to websites like Stack Overflow and follow things blindly, but using ChatGPT, they can at least know what they were doing wrong, and where they could improve.} -[P12]

ChatGPT commonly came up as an effective learning assistant or a tutor. Participants talked about using it for \hlcyan{step-by-step explanations of questions, for example, programming questions,} which are not possible when referring to online resources for assistance in learning. More so, its conversational modality can aid in self-study sessions as you can ask it to explain \hlcyan{certain concepts in detail and ask follow-up questions}. ChatGPT is also being used to \hlcyan{generate practice questions on a topic}.

\quotedtext{For example, I was in a course and the course did not have any pre-made content/questions. So sir asked ChatGPT to give us a bunch of questions for that course. And I think though that was kinda cool and that is how it was used in a good way.} -[P12]

Participants mentioned that ChatGPT is great at \hlcyan{generating theoretical responses}. It's frequently used for \hlcyan{lengthy reports, projects, and assignment write-ups}. Some participants even used it for internship applications and resume content. Despite occasional inaccuracies, many considered it a valuable content-writing tool. A participant mentioned -

\quotedtext{I used it religiously for one of the humanities courses where we had to write a report. I wrote almost the entire thing with the help of ChatGPT. Other than that I have used it to write some codes, get help in writing emails, etc.} -[P10]

\subsection{Challenges with the use of ChatGPT}
\customheading{Usability and Reliability Challenges.} Through our interviews, we got a sense of the usability challenges that the participants face while using ChatGPT, and how these factors affect their overall perception of ChatGPT and their usage behaviors around it. The process of accessing ChatGPT seemed to be a barrier to the effective use of the model. Participants reported that the process of signing into the website, and the periodically repetitive nature of re-authenticating with their accounts hindered their smooth experience. A participant explained -

\quotedtext{It's not like quillbot\footnote{https://quillbot.com/}, you open it, do your work, and get out. I have to make a new account here, so it just encroaches on my brain - a new account.} -[P13]

ChatGPT did not seem very humanistic to some participants, and they felt like it was \textit{too programmed} to talk to naturally, especially when they require a specific answer. This was highlighted when participants mentioned that \hlcyan{ChatGPT felt repetitive in its answers for theoretical questions despite providing it with different prompts}. Many participants found using it to be out of their comfort zone, requiring explicit context-setting which was tedious. Some found it more time-consuming than a typical Google search, as they needed to \hlcyan{verify every answer it provided explicitly}.

\quotedtext{\hlcyan{One tedious task is to write prompts. If you have to tell something to a human, you say it in just one sentence, but here to get accurate results, you have to write it in four to five sentences to give prompts.}} -[P5]

When it comes to providing the proper prompts for ChatGPT, we found that participants were not satisfied with the responses they got for the \hlcyan{long, complex prompts} provided by them. This made them break down their prompts into smaller segments to extract \hlcyan{better results}, and in some cases have to intervene on a line-to-line basis for their prompts. Writing well-structured prompts to get the right answer seemed like a challenge to many participants, and meant that having the skill to set the right context before asking their question was a prerequisite to get good quality results.

\quotedtext{\hlcyan{You have to double-check everything. You do it carefully, line by line, and make sure you exactly understand what it's doing. And most of the time, 80\% of the time, you'll have to edit it a lot to get it to work.}} -[P9]

Participants reported that ChatGPT does not meet their \hlcyan{academic and ideation requirements for course projects} as it fails to provide them with creative suggestions. Furthermore, participants found it to be \hlcyan{unreliable for programming}, as the more they used it for their \hlcyan{courses and debugging}, the more they reported encountering inaccuracies. Some participants highlighted the fact that ChatGPT \hlcyan{does not handle mathematical problems effectively}, causing them to rely on other sources for accurate answers. 

\quotedtext{\hlcyan{It tends to generate code snippets that might work but are not optimized or could be improved. In such cases, I have to manually refine the code. Also, there are times when it fails to understand complex questions properly, leading to irrelevant or inaccurate responses.}} -[P14] 

\customheading{Learning and Ethical Challenges.} Majority of the participants stated that they view ChatGPT in academics as a \textit{double-edged sword}. Participants appreciated ChatGPT as an assistant but suggested limiting its role to avoid making the human mind passive. Over-reliance could lead to \hlcyan{mental sluggishness and hinder effective learning and the cognitive thinking required to reach solutions to complex questions}. While useful for quick insights, participants noted its limitations in facilitating deep learning. One participant mentioned that it undermines rote learning, impacting their information retention.

\quotedtext{\hlcyan{It's helping as well, but at the same time, it's reducing the human power, and human thinking ability. Like we always depend on chat GPT for new ideas. We are not running our own minds.}} -[P5]

Some participants stated that they find the use of ChatGPT in academics unethical. They reported that using the tool for their academic tasks feels like an unethical shortcut, which leads to a feeling of guilt, and is a blocker for human efforts and creativity.

\quotedtext{Sometimes I feel as though, I am cheating by learning through it. I mean sometimes, you know, it just feels wrong when it gets too easy?} -[P7]

\subsection{Perceptions and Recommendations}

Our findings indicate that conventional practices like Google search, textbooks, and peer assistance are favored over ChatGPT. Trust issues with ChatGPT's authenticity and personal comfort zone hesitations were observed. Some participants prioritized personal effort for practices like coding and gathering information themselves, deterring their interest in using ChatGPT as an academic aid. A participant mentioned ChatGPT's limitations compared to human creativity and efficiency, affecting their reliance on it.

\quotedtext{\hlcyan{I might use it, but I feel there are sources that I'm already comfortable with [...] I Google it. It's more accessible. I do stick to very traditional sites where I can trust the answers.}} -[P13]

While participants expressed inhibitions towards adopting ChatGPT as an \hlcyan{integral part of their CS academic workflow}, they seemed hopeful about its future positive impact on education. A participant referred to it as a \textit{revolution in education} if it's used rightly. ChatGPT is being seen as the future `Google search'. Various participants said that this change is inevitable and we should accept it with an open mind. 

\quotedtext{\hlcyan{You can't go about your education without using Google. You need it for everything. It's incorporated into the system now. I feel the same, that's what will happen with ChatGPT in the future.}} -[P9] 

Participants suggested enhancing ChatGPT and related services for improved user experience. Multi-modality input was a common theme. They emphasized the necessity of training for efficient prompting and optimal use. A confidence score feature to authenticate the validity of answers was proposed. Effective multilingual capabilities for inclusivity were also suggested. \hlcyan{Some participants mentioned that domain-specific versions of ChatGPT for various CS courses can help offer insightful and reliable information in specific fields}.

\quotedtext{\hlcyan{Maybe it can start providing some tutorials on it soon. Is it possible? If it is possible, then it's really good. Like when we just open ChatGPT, it can show a few options like learn this, learn that I will make you learn this.}} -[P2]

%% file: files/05-discussion.tex
\subsection{Utilizing Existing Capabilities to Support Learning}
Our findings reveal various user patterns and benefits from ChatGPT's academic usage. Students utilized it for tasks like \hlcyan{code generation, error correction, brainstorming, concept tutoring, feedback, and content creation}. These patterns offer the potential for engineers and designers to develop technologies using LLMs to enhance these applications. Some of these patterns have also been explored in prior work \cite{becker2023ProsAndCons, Malinka2023Security, Daun2023Software, Denny2023CopilotCS1, Finnie-Ansley2022CS1, wermelinger2023Copilot, Savelka2023MCQAndCode, Reeves2023Parsons, finnie-ansley2023CodexCS2, Ouh2023Java, Cipriano2023GPT-3OOP, sarsa2022AutoGenerate, Leinonen2023CodeExplanation, Leinonen2023ExplainError, MacNeil2023CodeExplain, Balse2023Feedback}.

\customheading{Utilizing LLMs as Supervisors for Programming.} Our findings highlight the role of ChatGPT in providing programming assistance to students. Existing AI services such as GitHub Copilot \cite{Github_copilot} and OpenAI Codex \cite{OpenAI_Codex} currently assist coders with features like code completion and suggestions. However, our findings show that students appreciate a sense of control over their coding process and feel the need to understand and generate code in steps while testing their own expertise. Moreover, prior literature has shown that utilizing directly generated prompts can harm the learning experience of students \cite{becker2023ProsAndCons}. Therefore, we recommend frameworks and theories that break down the learning process into steps \cite{miner_using_2015, akella_learning_2010} to be incorporated in the design of LLM-powered agents developed to support students in practices like programming.

\customheading{Utilizing LLMs as Brainstorming Companions.} \hlcyan{Brainstorming and idea generation is another commonly reported use case of ChatGPT as shown in our findings}. We believe the capability of LLMs to \hlcyan{generate ideas and provide feedback} can be better leveraged in supporting the creativity of students while ensuring enough creative control and freedom for the students. Various studies have explored the supporting role of collaboration in ideation and brainstorming \cite{lacher_using_2018, lin_how_2021}. Assistants built on LLMs that can mimic human nature through texts can further support \hlcyan{online and offline collaboration sessions} among students. As shown from our findings, these assistants can be fine-tuned for \hlcyan{domain-specific assistance}. Abilities like multi-lingualism and multi-modality can further improve the quality of collaboration as shown in our findings.

\customheading{Utilizing LLMs as Tutors.} Our findings highlight the potential of LLMs to assist students like a tutor. Various studies have previously explored how AI-assisted chatbots can tutor students \cite{Denny2023CopilotCS1, leite_effects_2020}. An LLM can enhance this process owing to its higher knowledge base and human-like conversational interactions as reported by our participants. Our findings highlight \hlcyan{doubt solving, error detection, generation of practice questions and building of conceptual basics} as commonly reported uses of ChatGPT. These use cases can be used to build conversational tutors powered by LLMs. Our findings also highlight the \hlcyan{creation of subject-specific modes of LLMs} that can aid in a deeper understanding of specific subjects.

\subsection{Enhancing Utilization with Added Capabilities}
Our findings highlight various opportunities for improvement in LLM-based chatbots like ChatGPT. We discuss how these opportunities can be leveraged to create better learning experiences for students utilizing these agents. Concerns about the reliability and accuracy of ChatGPT responses were commonly highlighted in our findings. With the increasing research on LLMs \cite{liu_visual_2023, kong_better_2023}, the accuracy of these agents is bound to improve. However, it remains essential to inform the user about the accuracy of generated answers to protect them from potential harm caused by fake facts or hallucinations. Incorporation of features like confidence scores \cite{rechkemmer_when_2022} of the answers generated can help students stay away from incorrect information. 

More so, participants talked about ChatGPT seeming too programmed at times due to incidents like \hlcyan{repetition in its responses}. There is a need to enhance its \hlcyan{conversational abilities}. Providing features to report such incidents and customizing its mode and style of conversation can provide students more control over their interactions with ChatGPT. Such features can also help students to \hlcyan{fine-tune the content style of generated responses as per their requirements} as can be observed in Bing \cite{bing}. Concerns about the need to log in and create a new account to use ChatGPT also seemed to hinder the user experience of students. User workflows for minimal steps required to interact with the AI can support students with a simpler user journey. 

Concerns about possible plagiarism caused by some unethical users also commonly came up. Participants talked about the potential harm of ChatGPT to effective learning. It remains essential to ensure that \hlcyan{learning applications built on LLMs provide step-by-step learning to students and break solutions down into segments} that can better explain the solution. Incorporation of features like short quizzes and tests in such applications can ensure student learning while information provision. The need for training to better leverage ChatGPT also came up as a major finding. \hlcyan{Participants talked about their inability to come up with effective prompts to obtain preferred responses}. The creation of training guides that can guide students to rightly and ethically make use of ChatGPT can help educate students about the techniques and interactions that can improve their user experience. Upcoming prompting methods like in-context learning \cite{dong_survey_2023}, prompt engineering\cite{white_prompt_2023}, and chain of thought instructing\cite{wei_chain--thought_2023} in LLMs can be taught to students to help them obtain better results. ChatGPT like LLMs can also incorporate features to provide alternative prompts that match the intentions of the user better.

%% file: files/06-conclusion.tex
\hl{Large Language Models (LLMs) such as ChatGPT have become publicly accessible in the last one year or so and have garnered significant attention due to their potential impact across various domains. In this paper, we conduct a user study to analyze how ChatGPT is used by undergraduate computer science students in India. We find that students are actively using ChatGPT for tasks like generating and debugging code, brainstorming on new ideas, learning new concepts, getting feedback on their own solutions as well and creating new content such as reports and emails. Nevertheless, there are concerns regarding the reliability and accuracy of the responses generated by ChatGPT. These concerns could potentially be addressed by improving ChatGPT to provide confidence scores, data sources or additional details about its response generation process. Moreover, students themselves acknowledge that excessive reliance on ChatGPT could impede their long-term learning and personal development. Finally, we came across instances where students are unethically using ChatGPT to seek assistance in non-proctored assignments and examinations. 

Using the above findings, we propose a discussion on the development of various applications based on LLMs aimed at assisting students in their educational journey. Moreover, we discuss several factors that can aid in ensuring the broad adoption of these applications while mitigating any potential detrimental impacts. Looking ahead, we aim to delve deeper into the diverse use cases of these LLM-based applications for students. Our future research will involve scientifically evaluating the perceived advantages and disadvantages of these applications through in-classroom experiments and controlled experiments outside the classroom.} 

%% file: files/061-acknowledgements.tex
We are grateful to the anonymous reviewers for their feedback, which enhanced the overall quality of the paper. We thank Shamik Sinha, Palak Bhardwaj, Rayyan Hussain, Kashvi Panvanda, Sidhartha Garg, Shagun Yadav, Yash Chillar, and Sourav Goyal for their assistance in the development of questionnaires, as well as their involvement in conducting selected interviews.

%% file: files/07-appendix.tex
\vspace{-1em}
\section{Survey Questionnaire}\label{sec:survey_questionnaire}
Questions with standard bullets ($\bullet$) are single-select while questions with square ($\tiny\square$) are multi-select.
\begin{enumerate}
    \item What is your institution's name?
    \item In which year did you join the above-mentioned institute?

    $
        \bullet  \ 2018 \ \ \bullet  \ 2019 \ \ \bullet \ 2020 \ \ \bullet \ 2021 \ \bullet \ 2022 \ \ \bullet \ \textnormal{Others [Text Field]} $

    \item What is your degree program name?
    \item Are you familiar with ChatGPT, the AI assistive chatbot?
    $
        \bullet \ \textnormal{Very Familiar} \ \ \ \ \
        \bullet \ \textnormal{Somewhat Familiar} \ \ \ \ \
        \bullet \ \textnormal{Not Familiar}
        $
    
    \item How often do you use ChatGPT?
    $
        \bullet \ \textnormal{Daily} \ \ \
        \bullet \ \textnormal{Weekly} \ \ \
        \bullet \ \textnormal{Occasionally} \ \ \
        \bullet \ \textnormal{Rarely} \ \ \
        \bullet \ \textnormal{Never}
    $
    \item Since when have you been using ChatGPT? 
    \\
    $
        \bullet \ \textnormal{More than 3 months} \ \ \ \ \ \ \
        \bullet \ \textnormal{2-3 months} \\
        \bullet \ \textnormal{1-2 months} \ \ \ \ \ \ \ \ \ \ \ \ \ \ \ \ \ \ \ \
        \bullet \ \textnormal{Less than a month}
    $
    \renewcommand{\labelitemi}{\tiny$\square$}
    \item What have you used ChatGPT for in the past?
    \begin{itemize}
        \item Gathering information
        \item Summarizing content
        \item Assistance in essay writing
        \item Assistance in coding
        \item Assistance in assignments
        \item Assistance in generating content (emails, statements, etc.)
        \item Feedback and review
        \item Testing out the AI responses
        \item Other [Text field]
    \end{itemize}
    \item To what extent do you believe ChatGPT can aid in your coursework-related queries? Please select a rating from 1 to 5 where 1 indicates "Not useful at all" while 5 indicates "Highly Useful".
    \renewcommand{\labelitemi}{$\bullet$}
    $
        \bullet \ 1 \ \ \ \ \ \ \ \ \ \ \ \ \ \ \ 
        \bullet \ 2 \ \ \ \ \ \ \ \ \ \ \ \ \ \ \ 
        \bullet \ 3 \ \ \ \ \ \ \ \ \ \ \ \ \ \ \ 
        \bullet \ 4 \ \ \ \ \ \ \ \ \ \ \ \ \ \ \ 
        \bullet \ 5
    $
    \item How often do you run into problems using chatGPT? Note: Problems here refer to the issues in the responses generated by ChatGPT, such as false information, misunderstanding prompts, etc.
    $
        \bullet \ \textnormal{Never} \ \ \
        \bullet \ \textnormal{Rarely} \ \ \
        \bullet \ \textnormal{Sometimes} \ \ \
        \bullet \ \textnormal{Often} \ \ \
        \bullet \ \textnormal{Very often}
    $
    \item What are the potential challenges you foresee in integrating ChatGPT effectively into the learning and education process? You may choose NONE if you've never encountered problems.
    \renewcommand{\labelitemi}{\tiny$\square$}
    \begin{itemize}
        \item Incorrect and Unreliable answers
        \item Interface is inefficient
        \item It is not conversational enough
        \item I do not know the sources of information
        \item No way to verify the correctness
        \item Lack of trust
        \item None
    \end{itemize}
    \item What are the potential advantages you foresee in integrating ChatGPT effectively into the learning and education process? You may choose NONE if you do not see any advantages.
    \begin{itemize}
        \item Personalized Learning
        \item Immediate Feedback
        \item On-Demand Tutoring
        \item Engaging and Interactive Learning
        \item Access to More Information
        \item Continuous Assessment
        \item None
    \end{itemize}
    \item How would you describe your overall perception of ChatGPT as a learning and education tool?
    \renewcommand{\labelitemi}{$\bullet$}
    $
        \bullet \ \textnormal{Highly Positive} \ \ \ \ \
        \bullet \ \textnormal{Positive} \ \ \ \ \
        \bullet \ \textnormal{Neutral} \\
        \bullet \ \textnormal{Negative} \ \ \ \ \ \ \ \ \ \ \ \ \ \ \
        \bullet \ \textnormal{Highly Negative}
    $
    \item Can you elaborate your usage of ChatGPT? [Open-ended question. Participants were allowed to write without any word limit.]
    \item What are some suggestions you would want ChatGPT to incorporate, to make the experience better for you in an academic setting? [Open-ended question. Participants were allowed to write without any word limit.]
\end{enumerate}
\vspace{-1em}
\section{Interview Questionnaire}\label{sec:interview_questionnaire}
Below, we present the questionnaire template utilized by our research team during student interviews. It should be noted that this list does not cover all the questions, as the research team frequently posed additional inquiries in response to the participants' answers.
\begin{enumerate}
    \item How long have you been using ChatGPT for? How did you come across ChatGPT? How often do you use ChatGPT?
    \item What have you been using ChatGPT for? Do you have any specific use cases? Have you used ChatGPT before to learn new concepts?
    \item How do you think ChatGPT has supported your workflow, more specifically in your coursework?
    \item Have you run into problems with ChatGPT? Do you come across scenarios that might be undesirable or obstructing? How do you tackle such issues? How do you deal with cases that are not answered correctly by ChatGPT?
    \item What is your overall experience been like with ChatGPT? What is your current opinion of ChatGPT and the role it could play in changing the form of education? Threat or assistance?
    \item Are there any suggestions that you would like to incorporate in ChatGPT in order to support your computer science related workflow better?
\end{enumerate}

%% file: main.bbl

\begin{thebibliography}{43}


\ifx \showCODEN    \undefined \def \showCODEN     #1{\unskip}     \fi
\ifx \showDOI      \undefined \def \showDOI       #1{#1}\fi
\ifx \showISBNx    \undefined \def \showISBNx     #1{\unskip}     \fi
\ifx \showISBNxiii \undefined \def \showISBNxiii  #1{\unskip}     \fi
\ifx \showISSN     \undefined \def \showISSN      #1{\unskip}     \fi
\ifx \showLCCN     \undefined \def \showLCCN      #1{\unskip}     \fi
\ifx \shownote     \undefined \def \shownote      #1{#1}          \fi
\ifx \showarticletitle \undefined \def \showarticletitle #1{#1}   \fi
\ifx \showURL      \undefined \def \showURL       {\relax}        \fi
\providecommand\bibfield[2]{#2}
\providecommand\bibinfo[2]{#2}
\providecommand\natexlab[1]{#1}
\providecommand\showeprint[2][]{arXiv:#2}

\bibitem[alp({[n.\,d.]})]%
        {alphacode}
 \bibinfo{year}{[n.\,d.]}\natexlab{}.
\newblock \bibinfo{title}{AlphaCode}.
\newblock \bibinfo{howpublished}{\url{https://alphacode.deepmind.com/}}.
\newblock
\newblock
\shownote{Accessed: 2023-08-18}.


\bibitem[cod({[n.\,d.]})]%
        {codeWhisperer}
 \bibinfo{year}{[n.\,d.]}\natexlab{}.
\newblock \bibinfo{title}{Amazon CodeWhisperer}.
\newblock \bibinfo{howpublished}{\url{https://aws.amazon.com/codewhisperer/}}.
\newblock
\newblock
\shownote{Accessed: 2023-08-18}.


\bibitem[bin({[n.\,d.]})]%
        {bing}
 \bibinfo{year}{[n.\,d.]}\natexlab{}.
\newblock \bibinfo{title}{Bing}.
\newblock \bibinfo{howpublished}{\url{https://www.bing.com/}}.
\newblock
\newblock
\shownote{Accessed: 2023-08-18}.


\bibitem[Cha({[n.\,d.]})]%
        {ChatGPT}
 \bibinfo{year}{[n.\,d.]}\natexlab{}.
\newblock \bibinfo{title}{Chat GPT}.
\newblock \bibinfo{howpublished}{\url{https://chat.openai.com/}}.
\newblock
\newblock
\shownote{Accessed: 2023-08-18}.


\bibitem[Git({[n.\,d.]})]%
        {Github_copilot}
 \bibinfo{year}{[n.\,d.]}\natexlab{}.
\newblock \bibinfo{title}{Github Copilot}.
\newblock \bibinfo{howpublished}{\url{https://github.com/features/copilot}}.
\newblock
\newblock
\shownote{Accessed: 2023-08-18}.


\bibitem[Goo({[n.\,d.]})]%
        {Google_Bard}
 \bibinfo{year}{[n.\,d.]}\natexlab{}.
\newblock \bibinfo{title}{Google Bard}.
\newblock \bibinfo{howpublished}{\url{https://bard.google.com/}}.
\newblock
\newblock
\shownote{Accessed: 2023-08-18}.


\bibitem[gpt({[n.\,d.]})]%
        {gpt3}
 \bibinfo{year}{[n.\,d.]}\natexlab{}.
\newblock \bibinfo{title}{GPT-3}.
\newblock \bibinfo{howpublished}{\url{https://openai.com/blog/gpt-3-apps}}.
\newblock
\newblock
\shownote{Accessed: 2023-08-18}.


\bibitem[mix({[n.\,d.]})]%
        {mixed_methods}
 \bibinfo{year}{[n.\,d.]}\natexlab{}.
\newblock \bibinfo{title}{Mixed Methods Approach}.
\newblock \bibinfo{howpublished}{\url{https://en.wikipedia.org/wiki/Multimethodology}}.
\newblock
\newblock
\shownote{Accessed: 2023-08-18}.


\bibitem[Ope({[n.\,d.]})]%
        {OpenAI_Codex}
 \bibinfo{year}{[n.\,d.]}\natexlab{}.
\newblock \bibinfo{title}{OpenAI Codex}.
\newblock \bibinfo{howpublished}{\url{https://openai.com/blog/openai-codex}}.
\newblock
\newblock
\shownote{Accessed: 2023-08-18}.


\bibitem[Sno({[n.\,d.]})]%
        {Snowball}
 \bibinfo{year}{[n.\,d.]}\natexlab{}.
\newblock \bibinfo{title}{Snowball sampling}.
\newblock \bibinfo{howpublished}{\url{https://en.wikipedia.org/wiki/Snowball_sampling}}.
\newblock
\newblock
\shownote{Accessed: 2023-08-18}.


\bibitem[sta({[n.\,d.]})]%
        {stack_overflow}
 \bibinfo{year}{[n.\,d.]}\natexlab{}.
\newblock \bibinfo{title}{Stack Overflow}.
\newblock \bibinfo{howpublished}{\url{https://stackoverflow.com/}}.
\newblock
\newblock
\shownote{Accessed: 2023-08-18}.


\bibitem[Akella(2010)]%
        {akella_learning_2010}
\bibfield{author}{\bibinfo{person}{Devi Akella}.} \bibinfo{year}{2010}\natexlab{}.
\newblock \showarticletitle{Learning together: {Kolb}'s experiential theory and its application}.
\newblock \bibinfo{journal}{\emph{Journal of Management Organization - J MANAG ORGAN}}  \bibinfo{volume}{16} (\bibinfo{date}{March} \bibinfo{year}{2010}), \bibinfo{pages}{100--112}.
\newblock
\urldef\tempurl%
\url{https://doi.org/10.5172/jmo.16.1.100}
\showDOI{\tempurl}


\bibitem[Balse et~al\mbox{.}(2023)]%
        {Balse2023Feedback}
\bibfield{author}{\bibinfo{person}{Rishabh Balse}, \bibinfo{person}{Bharath Valaboju}, \bibinfo{person}{Shreya Singhal}, \bibinfo{person}{Jayakrishnan~Madathil Warriem}, {and} \bibinfo{person}{Prajish Prasad}.} \bibinfo{year}{2023}\natexlab{}.
\newblock \showarticletitle{Investigating the Potential of GPT-3 in Providing Feedback for Programming Assessments}. In \bibinfo{booktitle}{\emph{Proceedings of the ITiCSE 2023 Conference V. 1}} (Turku, Finland) \emph{(\bibinfo{series}{ITiCSE 2023})}. \bibinfo{publisher}{ACM}, \bibinfo{address}{New York, NY, USA}, \bibinfo{pages}{292–298}.
\newblock
\showISBNx{9798400701382}
\urldef\tempurl%
\url{https://doi.org/10.1145/3587102.3588852}
\showDOI{\tempurl}


\bibitem[Becker et~al\mbox{.}(2023)]%
        {becker2023ProsAndCons}
\bibfield{author}{\bibinfo{person}{Brett~A. Becker}, \bibinfo{person}{Paul Denny}, \bibinfo{person}{James Finnie-Ansley}, \bibinfo{person}{Andrew Luxton-Reilly}, \bibinfo{person}{James Prather}, {and} \bibinfo{person}{Eddie~Antonio Santos}.} \bibinfo{year}{2023}\natexlab{}.
\newblock \showarticletitle{Programming Is Hard - Or at Least It Used to Be: Educational Opportunities and Challenges of AI Code Generation}. In \bibinfo{booktitle}{\emph{Proceedings of the 54th ACM Technical Symposium on Computer Science Education V. 1}} (Toronto ON, Canada) \emph{(\bibinfo{series}{SIGCSE 2023})}. \bibinfo{publisher}{ACM}, \bibinfo{address}{New York, NY, USA}, \bibinfo{pages}{500–506}.
\newblock
\showISBNx{9781450394314}
\urldef\tempurl%
\url{https://doi.org/10.1145/3545945.3569759}
\showDOI{\tempurl}


\bibitem[Braun and Clarke(2006)]%
        {braun_using_2006}
\bibfield{author}{\bibinfo{person}{Virginia Braun} {and} \bibinfo{person}{Victoria Clarke}.} \bibinfo{year}{2006}\natexlab{}.
\newblock \showarticletitle{Using thematic analysis in psychology}.
\newblock \bibinfo{journal}{\emph{Qualitative Research in Psychology}}  \bibinfo{volume}{3} (\bibinfo{date}{Jan.} \bibinfo{year}{2006}), \bibinfo{pages}{77--101}.
\newblock
\urldef\tempurl%
\url{https://doi.org/10.1191/1478088706qp063oa}
\showDOI{\tempurl}


\bibitem[Cipriano and Alves(2023)]%
        {Cipriano2023GPT-3OOP}
\bibfield{author}{\bibinfo{person}{Bruno~Pereira Cipriano} {and} \bibinfo{person}{Pedro Alves}.} \bibinfo{year}{2023}\natexlab{}.
\newblock \showarticletitle{GPT-3 vs Object Oriented Programming Assignments: An Experience Report}. In \bibinfo{booktitle}{\emph{Proceedings of the 2023 ITiCSE Conference V. 1}} (Turku, Finland) \emph{(\bibinfo{series}{ITiCSE 2023})}. \bibinfo{publisher}{ACM}, \bibinfo{address}{New York, NY, USA}, \bibinfo{pages}{61–67}.
\newblock
\showISBNx{9798400701382}
\urldef\tempurl%
\url{https://doi.org/10.1145/3587102.3588814}
\showDOI{\tempurl}


\bibitem[Daun and Brings(2023)]%
        {Daun2023Software}
\bibfield{author}{\bibinfo{person}{Marian Daun} {and} \bibinfo{person}{Jennifer Brings}.} \bibinfo{year}{2023}\natexlab{}.
\newblock \showarticletitle{How ChatGPT Will Change Software Engineering Education}. In \bibinfo{booktitle}{\emph{Proceedings of the ITiCSE 2023 Conference V. 1}} (Turku, Finland) \emph{(\bibinfo{series}{ITiCSE 2023})}. \bibinfo{publisher}{ACM}, \bibinfo{address}{New York, NY, USA}, \bibinfo{pages}{110–116}.
\newblock
\showISBNx{9798400701382}
\urldef\tempurl%
\url{https://doi.org/10.1145/3587102.3588815}
\showDOI{\tempurl}


\bibitem[Denny et~al\mbox{.}(2023)]%
        {Denny2023CopilotCS1}
\bibfield{author}{\bibinfo{person}{Paul Denny}, \bibinfo{person}{Viraj Kumar}, {and} \bibinfo{person}{Nasser Giacaman}.} \bibinfo{year}{2023}\natexlab{}.
\newblock \showarticletitle{Conversing with Copilot: Exploring Prompt Engineering for Solving CS1 Problems Using Natural Language}. In \bibinfo{booktitle}{\emph{Proceedings of the 54th ACM Technical Symposium on Computer Science Education V. 1}} (Toronto ON, Canada) \emph{(\bibinfo{series}{SIGCSE 2023})}. \bibinfo{publisher}{ACM}, \bibinfo{address}{New York, NY, USA}, \bibinfo{pages}{1136–1142}.
\newblock
\showISBNx{9781450394314}
\urldef\tempurl%
\url{https://doi.org/10.1145/3545945.3569823}
\showDOI{\tempurl}


\bibitem[Dong et~al\mbox{.}(2023)]%
        {dong_survey_2023}
\bibfield{author}{\bibinfo{person}{Qingxiu Dong}, \bibinfo{person}{Lei Li}, \bibinfo{person}{Damai Dai}, \bibinfo{person}{Ce Zheng}, \bibinfo{person}{Zhiyong Wu}, \bibinfo{person}{Baobao Chang}, \bibinfo{person}{Xu Sun}, \bibinfo{person}{Jingjing Xu}, \bibinfo{person}{Lei Li}, {and} \bibinfo{person}{Zhifang Sui}.} \bibinfo{year}{2023}\natexlab{}.
\newblock \bibinfo{title}{A {Survey} on {In}-context {Learning}}.
\newblock
\newblock
\urldef\tempurl%
\url{https://doi.org/10.48550/arXiv.2301.00234}
\showDOI{\tempurl}
\newblock
\shownote{arXiv:2301.00234 [cs]}.


\bibitem[Finnie-Ansley et~al\mbox{.}(2022)]%
        {Finnie-Ansley2022CS1}
\bibfield{author}{\bibinfo{person}{James Finnie-Ansley}, \bibinfo{person}{Paul Denny}, \bibinfo{person}{Brett~A. Becker}, \bibinfo{person}{Andrew Luxton-Reilly}, {and} \bibinfo{person}{James Prather}.} \bibinfo{year}{2022}\natexlab{}.
\newblock \showarticletitle{The Robots Are Coming: Exploring the Implications of OpenAI Codex on Introductory Programming}. In \bibinfo{booktitle}{\emph{Proceedings of the 24th Australasian Computing Education Conference}} (Virtual Event, Australia) \emph{(\bibinfo{series}{ACE '22})}. \bibinfo{publisher}{ACM}, \bibinfo{address}{New York, NY, USA}, \bibinfo{pages}{10–19}.
\newblock
\showISBNx{9781450396431}
\urldef\tempurl%
\url{https://doi.org/10.1145/3511861.3511863}
\showDOI{\tempurl}


\bibitem[Finnie-Ansley et~al\mbox{.}(2023)]%
        {finnie-ansley2023CodexCS2}
\bibfield{author}{\bibinfo{person}{James Finnie-Ansley}, \bibinfo{person}{Paul Denny}, \bibinfo{person}{Andrew Luxton-Reilly}, \bibinfo{person}{Eddie~Antonio Santos}, \bibinfo{person}{James Prather}, {and} \bibinfo{person}{Brett~A. Becker}.} \bibinfo{year}{2023}\natexlab{}.
\newblock \showarticletitle{My AI Wants to Know If This Will Be on the Exam: Testing OpenAI’s Codex on CS2 Programming Exercises}. In \bibinfo{booktitle}{\emph{Proceedings of the 25th ACE Conference}} (Melbourne, VIC, Australia) \emph{(\bibinfo{series}{ACE '23})}. \bibinfo{publisher}{ACM}, \bibinfo{address}{New York, NY, USA}, \bibinfo{pages}{97–104}.
\newblock
\showISBNx{9781450399418}
\urldef\tempurl%
\url{https://doi.org/10.1145/3576123.3576134}
\showDOI{\tempurl}


\bibitem[Firat(2023)]%
        {Mehmet23studentperception}
\bibfield{author}{\bibinfo{person}{Mehmet Firat}.} \bibinfo{year}{2023}\natexlab{}.
\newblock \showarticletitle{What ChatGPT means for universities: Perceptions of scholars and students}.
\newblock \bibinfo{journal}{\emph{Journal of Applied Learning \& Teaching}} \bibinfo{volume}{6}, \bibinfo{number}{1} (\bibinfo{year}{2023}), \bibinfo{pages}{1--22}.
\newblock
\urldef\tempurl%
\url{https://doi.org/10.37074/jalt.2023.6.1.22}
\showDOI{\tempurl}


\bibitem[Floridi and Chiriatti(2020)]%
        {floridi2020gpt}
\bibfield{author}{\bibinfo{person}{Luciano Floridi} {and} \bibinfo{person}{Massimo Chiriatti}.} \bibinfo{year}{2020}\natexlab{}.
\newblock \showarticletitle{GPT-3: Its nature, scope, limits, and consequences}.
\newblock \bibinfo{journal}{\emph{Minds and Machines}}  \bibinfo{volume}{30} (\bibinfo{year}{2020}), \bibinfo{pages}{681--694}.
\newblock


\bibitem[Kong et~al\mbox{.}(2023)]%
        {kong_better_2023}
\bibfield{author}{\bibinfo{person}{Aobo Kong}, \bibinfo{person}{Shiwan Zhao}, \bibinfo{person}{Hao Chen}, \bibinfo{person}{Qicheng Li}, \bibinfo{person}{Yong Qin}, \bibinfo{person}{Ruiqi Sun}, {and} \bibinfo{person}{Xin Zhou}.} \bibinfo{year}{2023}\natexlab{}.
\newblock \bibinfo{title}{Better {Zero}-{Shot} {Reasoning} with {Role}-{Play} {Prompting}}.
\newblock
\newblock
\urldef\tempurl%
\url{https://doi.org/10.48550/arXiv.2308.07702}
\showDOI{\tempurl}
\newblock
\shownote{arXiv:2308.07702 [cs]}.


\bibitem[Lacher and Biehl(2018)]%
        {lacher_using_2018}
\bibfield{author}{\bibinfo{person}{Lisa Lacher} {and} \bibinfo{person}{Cydnee Biehl}.} \bibinfo{year}{2018}\natexlab{}.
\newblock \showarticletitle{Using {Discord} to {Understand} and {Moderate} {Collaboration} and {Teamwork}: ({Abstract} {Only})}. In \bibinfo{booktitle}{\emph{Proceedings of the 49th {ACM} {Technical} {Symposium} on {Computer} {Science} {Education}}} \emph{(\bibinfo{series}{{SIGCSE} '18})}. \bibinfo{publisher}{Association for Computing Machinery}, \bibinfo{address}{New York, NY, USA}, \bibinfo{pages}{1107}.
\newblock
\showISBNx{978-1-4503-5103-4}
\urldef\tempurl%
\url{https://doi.org/10.1145/3159450.3162231}
\showDOI{\tempurl}


\bibitem[Leinonen et~al\mbox{.}(2023a)]%
        {Leinonen2023CodeExplanation}
\bibfield{author}{\bibinfo{person}{Juho Leinonen}, \bibinfo{person}{Paul Denny}, \bibinfo{person}{Stephen MacNeil}, \bibinfo{person}{Sami Sarsa}, \bibinfo{person}{Seth Bernstein}, \bibinfo{person}{Joanne Kim}, \bibinfo{person}{Andrew Tran}, {and} \bibinfo{person}{Arto Hellas}.} \bibinfo{year}{2023}\natexlab{a}.
\newblock \showarticletitle{Comparing Code Explanations Created by Students and Large Language Models}. In \bibinfo{booktitle}{\emph{Proceedings of the ITiCSE 2023 Conference V. 1}} (Turku, Finland) \emph{(\bibinfo{series}{ITiCSE 2023})}. \bibinfo{publisher}{ACM}, \bibinfo{address}{New York, NY, USA}, \bibinfo{pages}{124–130}.
\newblock
\showISBNx{9798400701382}
\urldef\tempurl%
\url{https://doi.org/10.1145/3587102.3588785}
\showDOI{\tempurl}


\bibitem[Leinonen et~al\mbox{.}(2023b)]%
        {Leinonen2023ExplainError}
\bibfield{author}{\bibinfo{person}{Juho Leinonen}, \bibinfo{person}{Arto Hellas}, \bibinfo{person}{Sami Sarsa}, \bibinfo{person}{Brent Reeves}, \bibinfo{person}{Paul Denny}, \bibinfo{person}{James Prather}, {and} \bibinfo{person}{Brett~A. Becker}.} \bibinfo{year}{2023}\natexlab{b}.
\newblock \showarticletitle{Using Large Language Models to Enhance Programming Error Messages}. In \bibinfo{booktitle}{\emph{Proceedings of the 54th ACM Technical Symposium on Computer Science Education V. 1}} (Toronto ON, Canada) \emph{(\bibinfo{series}{SIGCSE 2023})}. \bibinfo{publisher}{ACM}, \bibinfo{address}{New York, NY, USA}, \bibinfo{pages}{563–569}.
\newblock
\showISBNx{9781450394314}
\urldef\tempurl%
\url{https://doi.org/10.1145/3545945.3569770}
\showDOI{\tempurl}


\bibitem[Leite and Blanco(2020)]%
        {leite_effects_2020}
\bibfield{author}{\bibinfo{person}{Abe Leite} {and} \bibinfo{person}{Saúl~A. Blanco}.} \bibinfo{year}{2020}\natexlab{}.
\newblock \showarticletitle{Effects of {Human} vs. {Automatic} {Feedback} on {Students}' {Understanding} of {AI} {Concepts} and {Programming} {Style}}. In \bibinfo{booktitle}{\emph{Proceedings of the 51st {ACM} {Technical} {Symposium} on {Computer} {Science} {Education}}} \emph{(\bibinfo{series}{{SIGCSE} '20})}. \bibinfo{publisher}{ACM}, \bibinfo{address}{New York, NY, USA}, \bibinfo{pages}{44--50}.
\newblock
\showISBNx{978-1-4503-6793-6}
\urldef\tempurl%
\url{https://doi.org/10.1145/3328778.3366921}
\showDOI{\tempurl}


\bibitem[Lin et~al\mbox{.}(2021)]%
        {lin_how_2021}
\bibfield{author}{\bibinfo{person}{Xinyue Lin}, \bibinfo{person}{James Connors}, \bibinfo{person}{Chang Lim}, {and} \bibinfo{person}{John~R. Hott}.} \bibinfo{year}{2021}\natexlab{}.
\newblock \showarticletitle{How {Do} {Students} {Collaborate}? {Analyzing} {Group} {Choice} in a {Collaborative} {Learning} {Environment}}. In \bibinfo{booktitle}{\emph{Proceedings of the 52nd {ACM} {Technical} {Symposium} on {Computer} {Science} {Education}}} \emph{(\bibinfo{series}{{SIGCSE} '21})}. \bibinfo{publisher}{ACM}, \bibinfo{address}{New York, NY, USA}, \bibinfo{pages}{212--218}.
\newblock
\showISBNx{978-1-4503-8062-1}
\urldef\tempurl%
\url{https://doi.org/10.1145/3408877.3432389}
\showDOI{\tempurl}


\bibitem[Liu et~al\mbox{.}(2023)]%
        {liu_visual_2023}
\bibfield{author}{\bibinfo{person}{Haotian Liu}, \bibinfo{person}{Chunyuan Li}, \bibinfo{person}{Qingyang Wu}, {and} \bibinfo{person}{Yong~Jae Lee}.} \bibinfo{year}{2023}\natexlab{}.
\newblock \bibinfo{title}{Visual {Instruction} {Tuning}}.
\newblock
\newblock
\urldef\tempurl%
\url{https://doi.org/10.48550/arXiv.2304.08485}
\showDOI{\tempurl}
\newblock
\shownote{arXiv:2304.08485 [cs]}.


\bibitem[MacNeil et~al\mbox{.}(2023)]%
        {MacNeil2023CodeExplain}
\bibfield{author}{\bibinfo{person}{Stephen MacNeil}, \bibinfo{person}{Andrew Tran}, \bibinfo{person}{Arto Hellas}, \bibinfo{person}{Joanne Kim}, \bibinfo{person}{Sami Sarsa}, \bibinfo{person}{Paul Denny}, \bibinfo{person}{Seth Bernstein}, {and} \bibinfo{person}{Juho Leinonen}.} \bibinfo{year}{2023}\natexlab{}.
\newblock \showarticletitle{Experiences from Using Code Explanations Generated by Large Language Models in a Web Software Development E-Book}. In \bibinfo{booktitle}{\emph{Proceedings of the 54th ACM Technical Symposium on Computer Science Education V. 1}} (Toronto ON, Canada) \emph{(\bibinfo{series}{SIGCSE 2023})}. \bibinfo{publisher}{ACM}, \bibinfo{address}{New York, NY, USA}, \bibinfo{pages}{931–937}.
\newblock
\showISBNx{9781450394314}
\urldef\tempurl%
\url{https://doi.org/10.1145/3545945.3569785}
\showDOI{\tempurl}


\bibitem[Malinka et~al\mbox{.}(2023)]%
        {Malinka2023Security}
\bibfield{author}{\bibinfo{person}{Kamil Malinka}, \bibinfo{person}{Martin Peres\'{\i}ni}, \bibinfo{person}{Anton Firc}, \bibinfo{person}{Ondrej Hujn\'{a}k}, {and} \bibinfo{person}{Filip Janus}.} \bibinfo{year}{2023}\natexlab{}.
\newblock \showarticletitle{On the Educational Impact of ChatGPT: Is Artificial Intelligence Ready to Obtain a University Degree?}. In \bibinfo{booktitle}{\emph{Proceedings of the ITiCSE 2023 Conference V. 1}} (Turku, Finland) \emph{(\bibinfo{series}{ITiCSE 2023})}. \bibinfo{publisher}{ACM}, \bibinfo{address}{New York, NY, USA}, \bibinfo{pages}{47–53}.
\newblock
\showISBNx{9798400701382}
\urldef\tempurl%
\url{https://doi.org/10.1145/3587102.3588827}
\showDOI{\tempurl}


\bibitem[Miner et~al\mbox{.}(2015)]%
        {miner_using_2015}
\bibfield{author}{\bibinfo{person}{Amy Miner}, \bibinfo{person}{Jennifer Mallow}, \bibinfo{person}{Laurie Theeke}, {and} \bibinfo{person}{Emily Barnes}.} \bibinfo{year}{2015}\natexlab{}.
\newblock \showarticletitle{Using {Gagne}'s 9 {Events} of {Instruction} to {Enhance} {Student} {Performance} and {Course} {Evaluations} in {Undergraduate} {Nursing} {Course}}.
\newblock \bibinfo{journal}{\emph{Nurse educator}} \bibinfo{volume}{40}, \bibinfo{number}{3} (\bibinfo{year}{2015}), \bibinfo{pages}{152--154}.
\newblock
\showISSN{0363-3624}
\urldef\tempurl%
\url{https://doi.org/10.1097/NNE.0000000000000138}
\showDOI{\tempurl}


\bibitem[OpenAI(2016)]%
        {gen_ai_web_link}
\bibfield{author}{\bibinfo{person}{OpenAI}.} \bibinfo{year}{2016}\natexlab{}.
\newblock \bibinfo{title}{Generative models}.
\newblock
\newblock
\urldef\tempurl%
\url{https://openai.com/research/generative-models}
\showURL{%
\tempurl}


\bibitem[Ouh et~al\mbox{.}(2023)]%
        {Ouh2023Java}
\bibfield{author}{\bibinfo{person}{Eng~Lieh Ouh}, \bibinfo{person}{Benjamin Kok~Siew Gan}, \bibinfo{person}{Kyong Jin~Shim}, {and} \bibinfo{person}{Swavek Wlodkowski}.} \bibinfo{year}{2023}\natexlab{}.
\newblock \showarticletitle{ChatGPT, Can You Generate Solutions for My Coding Exercises? An Evaluation on Its Effectiveness in an Undergraduate Java Programming Course.}. In \bibinfo{booktitle}{\emph{Proceedings of the ITiCSE 2023 Conference V. 1}} (Turku, Finland) \emph{(\bibinfo{series}{ITiCSE 2023})}. \bibinfo{publisher}{ACM}, \bibinfo{address}{New York, NY, USA}, \bibinfo{pages}{54–60}.
\newblock
\showISBNx{9798400701382}
\urldef\tempurl%
\url{https://doi.org/10.1145/3587102.3588794}
\showDOI{\tempurl}


\bibitem[Rechkemmer and Yin(2022)]%
        {rechkemmer_when_2022}
\bibfield{author}{\bibinfo{person}{Amy Rechkemmer} {and} \bibinfo{person}{Ming Yin}.} \bibinfo{year}{2022}\natexlab{}.
\newblock \showarticletitle{When {Confidence} {Meets} {Accuracy}: {Exploring} the {Effects} of {Multiple} {Performance} {Indicators} on {Trust} in {Machine} {Learning} {Models}}. In \bibinfo{booktitle}{\emph{Proceedings of the 2022 {CHI} {Conference} on {Human} {Factors} in {Computing} {Systems}}} \emph{(\bibinfo{series}{{CHI} '22})}. \bibinfo{publisher}{ACM}, \bibinfo{address}{New York, NY, USA}, \bibinfo{pages}{1--14}.
\newblock
\showISBNx{978-1-4503-9157-3}
\urldef\tempurl%
\url{https://doi.org/10.1145/3491102.3501967}
\showDOI{\tempurl}


\bibitem[Reeves et~al\mbox{.}(2023)]%
        {Reeves2023Parsons}
\bibfield{author}{\bibinfo{person}{Brent Reeves}, \bibinfo{person}{Sami Sarsa}, \bibinfo{person}{James Prather}, \bibinfo{person}{Paul Denny}, \bibinfo{person}{Brett~A. Becker}, \bibinfo{person}{Arto Hellas}, \bibinfo{person}{Bailey Kimmel}, \bibinfo{person}{Garrett Powell}, {and} \bibinfo{person}{Juho Leinonen}.} \bibinfo{year}{2023}\natexlab{}.
\newblock \showarticletitle{Evaluating the Performance of Code Generation Models for Solving Parsons Problems With Small Prompt Variations}. In \bibinfo{booktitle}{\emph{Proceedings of the ITiCSE 2023 Conference V. 1}} (Turku, Finland) \emph{(\bibinfo{series}{ITiCSE 2023})}. \bibinfo{publisher}{ACM}, \bibinfo{address}{New York, NY, USA}, \bibinfo{pages}{299–305}.
\newblock
\showISBNx{9798400701382}
\urldef\tempurl%
\url{https://doi.org/10.1145/3587102.3588805}
\showDOI{\tempurl}


\bibitem[Sarsa et~al\mbox{.}(2022)]%
        {sarsa2022AutoGenerate}
\bibfield{author}{\bibinfo{person}{Sami Sarsa}, \bibinfo{person}{Paul Denny}, \bibinfo{person}{Arto Hellas}, {and} \bibinfo{person}{Juho Leinonen}.} \bibinfo{year}{2022}\natexlab{}.
\newblock \showarticletitle{Automatic Generation of Programming Exercises and Code Explanations Using Large Language Models}. In \bibinfo{booktitle}{\emph{Proceedings of the 2022 ACM Conference on International Computing Education Research - Volume 1}} (Lugano and Virtual Event, Switzerland) \emph{(\bibinfo{series}{ICER '22})}. \bibinfo{publisher}{ACM}, \bibinfo{address}{New York, NY, USA}, \bibinfo{pages}{27–43}.
\newblock
\showISBNx{9781450391948}
\urldef\tempurl%
\url{https://doi.org/10.1145/3501385.3543957}
\showDOI{\tempurl}


\bibitem[Savelka et~al\mbox{.}(2023)]%
        {Savelka2023MCQAndCode}
\bibfield{author}{\bibinfo{person}{Jaromir Savelka}, \bibinfo{person}{Arav Agarwal}, \bibinfo{person}{Christopher Bogart}, \bibinfo{person}{Yifan Song}, {and} \bibinfo{person}{Majd Sakr}.} \bibinfo{year}{2023}\natexlab{}.
\newblock \showarticletitle{Can Generative Pre-Trained Transformers (GPT) Pass Assessments in Higher Education Programming Courses?}. In \bibinfo{booktitle}{\emph{Proceedings of the 2023 ITiCSE Conference V. 1}} (Turku, Finland) \emph{(\bibinfo{series}{ITiCSE 2023})}. \bibinfo{publisher}{ACM}, \bibinfo{address}{New York, NY, USA}, \bibinfo{pages}{117–123}.
\newblock
\showISBNx{9798400701382}
\urldef\tempurl%
\url{https://doi.org/10.1145/3587102.3588792}
\showDOI{\tempurl}


\bibitem[Shoufan(2023)]%
        {Shoufan23studentperception}
\bibfield{author}{\bibinfo{person}{Abdulhadi Shoufan}.} \bibinfo{year}{2023}\natexlab{}.
\newblock \showarticletitle{Exploring Students’ Perceptions of ChatGPT: Thematic Analysis and Follow-Up Survey}.
\newblock \bibinfo{journal}{\emph{IEEE Access}}  \bibinfo{volume}{11} (\bibinfo{year}{2023}), \bibinfo{pages}{38805--38818}.
\newblock
\urldef\tempurl%
\url{https://doi.org/10.1109/ACCESS.2023.3268224}
\showDOI{\tempurl}


\bibitem[Wei et~al\mbox{.}(2023)]%
        {wei_chain--thought_2023}
\bibfield{author}{\bibinfo{person}{Jason Wei}, \bibinfo{person}{Xuezhi Wang}, \bibinfo{person}{Dale Schuurmans}, \bibinfo{person}{Maarten Bosma}, \bibinfo{person}{Brian Ichter}, \bibinfo{person}{Fei Xia}, \bibinfo{person}{Ed Chi}, \bibinfo{person}{Quoc Le}, {and} \bibinfo{person}{Denny Zhou}.} \bibinfo{year}{2023}\natexlab{}.
\newblock \bibinfo{title}{Chain-of-{Thought} {Prompting} {Elicits} {Reasoning} in {Large} {Language} {Models}}.
\newblock
\newblock
\urldef\tempurl%
\url{https://doi.org/10.48550/arXiv.2201.11903}
\showDOI{\tempurl}
\newblock
\shownote{arXiv:2201.11903 [cs]}.


\bibitem[Wermelinger(2023)]%
        {wermelinger2023Copilot}
\bibfield{author}{\bibinfo{person}{Michel Wermelinger}.} \bibinfo{year}{2023}\natexlab{}.
\newblock \showarticletitle{Using GitHub Copilot to Solve Simple Programming Problems}. In \bibinfo{booktitle}{\emph{Proceedings of the 54th ACM SIGCSE TS V. 1}} (Toronto ON, Canada) \emph{(\bibinfo{series}{SIGCSE 2023})}. \bibinfo{publisher}{ACM}, \bibinfo{address}{New York, NY, USA}, \bibinfo{pages}{172–178}.
\newblock
\showISBNx{9781450394314}
\urldef\tempurl%
\url{https://doi.org/10.1145/3545945.3569830}
\showDOI{\tempurl}


\bibitem[White et~al\mbox{.}(2023)]%
        {white_prompt_2023}
\bibfield{author}{\bibinfo{person}{Jules White}, \bibinfo{person}{Quchen Fu}, \bibinfo{person}{Sam Hays}, \bibinfo{person}{Michael Sandborn}, \bibinfo{person}{Carlos Olea}, \bibinfo{person}{Henry Gilbert}, \bibinfo{person}{Ashraf Elnashar}, \bibinfo{person}{Jesse Spencer-Smith}, {and} \bibinfo{person}{Douglas~C. Schmidt}.} \bibinfo{year}{2023}\natexlab{}.
\newblock \bibinfo{title}{A {Prompt} {Pattern} {Catalog} to {Enhance} {Prompt} {Engineering} with {ChatGPT}}.
\newblock
\newblock
\urldef\tempurl%
\url{https://doi.org/10.48550/arXiv.2302.11382}
\showDOI{\tempurl}
\newblock
\shownote{arXiv:2302.11382 [cs]}.


\end{thebibliography}
